\documentclass[aip,graphicx]{revtex4-2}

\usepackage{amsmath}
\usepackage{graphicx} 
\usepackage{subfig} 
\usepackage{amssymb}
\usepackage{trfsigns}
\usepackage{color}
\usepackage{algorithm}

\usepackage[usenames,dvipsnames]{xcolor}
\usepackage{dcolumn}
\usepackage{mathtools}
\usepackage{tikz}
\usepackage{appendix}
\usepackage{amsfonts}
\usepackage{amsmath}
\usepackage{amssymb}
\usepackage{amsthm}
\usepackage{float}
\usepackage{subfig}

\newcommand{\ds}{\displaystyle}

\floatname{algorithm}{Table}

\begin{document}


\title{On the use of asymptotically motivated gauge functions to obtain convergent series solutions to nonlinear ODEs}  



\author{Nastaran Naghshineh}
\email[corresponding author: ]{nxncad@rit.edu}
\affiliation{School of Mathematics and Statistics, Rochester Institute of Technology, Rochester, NY, 14623, USA}
\affiliation{Department of Mathematics and Sciences, Rochester Institute of Technology-Dubai, Dubai, 341055, UAE}

\author{W. Cade Reinberger}
\affiliation{School of Mathematics and Statistics, Rochester Institute of Technology, Rochester, NY, 14623, USA}

\author{Nathaniel S. Barlow}
\affiliation{School of Mathematics and Statistics, Rochester Institute of Technology, Rochester, NY, 14623, USA}

\author{Mohamed A. Samaha}
\affiliation{Department of Mechanical and Industrial Engineering, Rochester Institute of Technology-Dubai, Dubai, 341055, UAE}
\affiliation{School of Mathematics and Statistics, Rochester Institute of Technology, Rochester, NY, 14623, USA}
        
\author{Steven J. Weinstein}
\affiliation{School of Mathematics and Statistics, Rochester Institute of Technology, Rochester, NY, 14623, USA}

\affiliation{Department of Chemical Engineering, Rochester Institute of Technology, 
Rochester, NY, 14623, USA}



\date{\today}
\begin{abstract}
{We examine the power series solutions of two classical nonlinear ordinary differential equations of fluid mechanics that are mathematically related by their large-distance asymptotic behaviors in semi-infinite domains. The first problem is that of the ``Sakiadis'' boundary layer over a moving flat wall, for which no exact analytic solution has been put forward. The second problem is that of a static air--liquid meniscus with surface tension that intersects a flat wall at a given contact angle and limits to a flat pool away from the wall.  For the latter problem, the exact analytic solution---given as distance from the wall as function of meniscus height---has long been known (Batchelor, 1967). Here, we provide an explicit solution as meniscus height vs.~distance from the wall to elucidate structural similarities to the Sakiadis boundary layer.   Although power series solutions are readily obtainable to the governing nonlinear ordinary differential equations, we show that---in both problems---the series diverge due to non-physical complex or negative real-valued singularities. In both cases, these singularities can be moved by expanding in exponential gauge functions motivated by their respective large distance asymptotic behaviors to enable series convergence over their full semi-infinite domains. For the Sakiadis problem, this not only provides a convergent Taylor series (and conjectured exact) solution to the ODE, but also a means to evaluate the wall shear parameter (and other properties) to within any desired precision. Although the nature of nonlinear ODEs precludes general conclusions, our results indicate that asymptotic behaviors can be useful when proposing variable transformations to overcome power series divergence.}
\end{abstract}
\maketitle 

\section{Introduction \label{sec:intro}} 

Infinite power series solutions to ordinary differential equations (ODEs) are useful if they are convergent while also satisfying the given constraints over the domain on which they are defined. It is within this context that we develop a convergent series solution to a classical nonlinear ODE in fluid mechanics--the Sakiadis boundary layer flow along a moving wall.  To date, this problem, as well as the related well-known Blasius problem along a stationary wall, do not have proven exact analytical solutions in the literature, although approximate analytical solutions have been put forward and are discussed later in this section.  The Sakiadis boundary layer is an important flow field in configurations where thin liquid films are coated onto moving substrates~\citep{Weinstein2004}, and is an essential component of hydrodynamic assist in high speed curtain coating~\citep{Blake}.

The boundary layer surrounding a flat plate moving through a viscous incompressible fluid was first examined in the literature by~\citet{Sakiadis} who applied Blasius's similarity transform to Prandtl's boundary layer equations (with appropriate boundary conditions, B.C.s) to arrive at a third-order nonlinear ODE in $f(\eta)$ refered to here as the `Sakiadis Problem',
\begin{subequations}
\begin{equation}
2f'''+ff''=0,~~~0\le\eta<\infty
\label{sDE}
\end{equation}
\begin{equation}
\text{Sakiadis B.C.s:  }~f(0)=0, ~ f'(0)=1, ~ f'(\infty) = 0.
\label{sBC}
\end{equation}
\label{eq:Sakiadis}
\end{subequations}
By contrast, the Blasius problem describing a stationary plate in a moving fluid is governed by the same operator~(\ref{sDE}) but has conditions
\begin{equation}
\text{Blasius B.C.s:  }~f(0)=0, ~ f'(0)=0, ~ f'(\infty) = 1.
\label{bBC}
\end{equation}

While both the Sakiadis and Blasius problems can be handled in similar ways numerically (e.g., shooting, transformation)~\citep{Cebeci,Fazio1992,Cortell,Eftekhari,Fazio2015}, the difference in boundary conditions leads to different (approximate) analytical approaches \citep{Barlow:2017}; and, the nonlinear nature of the equations yields distinctly different solutions. A common measure of the accuracy of any solution technique applied to either problem is the quantity $\kappa$, defined as
\begin{equation}
    \kappa\equiv f''(0),
    \label{eq:kappa}
\end{equation}
which is directly related to the wall shear stress in the boundary-layer flow, and is typically referred to as the ``wall shear'' parameter~\citep{Cortell,Fazio2015}.  For a given $\kappa$---which is known numerically and is determined in this paper algorithmically---an infinite power series solution for both problems can be obtained through standard means, given as 
\begin{equation}
    f=\sum_{n=0}^\infty a_n \eta^n,~~|\eta|<S,
    \label{eq:series}
\end{equation}
where $S$ is a finite radius of convergence.  Singularities that lie outside of the physical domain at a distance $S$ away from $\eta=0$ are the cause of this radius. The Blasius problem has 3 well-known singularities lying the same distance $S$ from the origin (see~\citet{Boyd1999} and historical review therein) and are reported to lie at values $\eta=S\text{ exp }[i(2j+1)\pi/3]$ ($j$=0, 1, 2) where $S\approx 5.6900380545$.   For the Sakiadis problem,  $S\approx4.07217$, arising from singularities lying off the real-line in the left half-plane~\citep{Barlow:2017}.  Approximate resummations are available, that bypass the original series' convergence barrier caused by these singularities for both the Sakiadis~\citep{Barlow:2017} and Blasius~\citep{Boyd1997,Boyd1999,Abbasbandy,Barlow:2017} problems. 


One approach for avoiding (non-physical) singularities that restrict series convergence is to re-expand the series by mapping the independent variable such that the (non-physical) closest singularities no longer affect the physical domain. The divergent Blasius series has been successfully re-summed in this way by ~\citet{Boyd1999} through re-expanding~(\ref{eq:series}) as
\begin{subequations}
\begin{equation}
    \text{Blasius:  }f=\sum_{n=0}^\infty \tilde{b}_n \left[\delta(\eta)\right]^n,~~|\delta(\eta)|<\tilde{S},
    \label{eq:Boydseries}
\end{equation}
where the expansion variable
\begin{equation}
    \delta(\eta)=\frac{2\eta^3}{S^3+\eta^3}
    \label{eq:gauge}
\end{equation}
    \label{eq:Boyd}
\end{subequations}is the \textit{gauge function} by which~(\ref{eq:Boydseries}) may be cast as a formal Taylor series, specifying how terms are asymptotically ordered~\citep{VanDykeChapterGauge,Leal}.  In~(\ref{eq:Boyd}), the three-fold symmetry of the closest singularities (with modulus $S$) in the complex $\eta$-plane are mapped to infinity, due to the effect of the above transformation on their orientation.  This leads to a new radius of convergence $\tilde{S}$ that lies outside the original physical domain, thus creating a convergent series solution for the Blasius problem on the positive real line. The coefficients $\tilde{b}_n$ are obtained by equating ~(\ref{eq:series}) with the expansion of~(\ref{eq:Boyd}) about $\eta=0$; these may be obtained recursively (only depending on prior coefficients) due to the gauge function~(\ref{eq:gauge}) and its derivatives equaling $0$ at $\eta=0$.  The substitution made in~(\ref{eq:Boyd}) may be considered a modified Euler transformation\footnote{There are several definitions of Euler transformations in the literature. The definitions and interpretations that align most with this work are found in~\cite{VanDykeChapter,BakerCritical}}.  

While the series~(\ref{eq:Boyd}) explicitly incorporates singularities in order to bypass the radius of convergence of the original series~(\ref{eq:series}), another approach is to consider the other side of the domain ($\eta\to\infty$) as a possible expansion point.  Using the method of dominant balance, the $\eta\to\infty$ behaviors for the Blasius and Sakiadis solutions are given in~\cite{Barlow:2017} as 
\begin{equation}
\text{Blasius:  }f \sim \eta+B+4Q  \frac{\exp[-\eta^2/4-B\eta/2]}{(\eta+B)^2}\left[1+O\left(\frac{1}{(\eta+B)^2}\right)\right],~\textrm{as} \ \eta \rightarrow \infty,
\label{eq:blasiusA}
\end{equation}
\begin{equation}
\text{Sakiadis:  }f \sim C+ G e^{-C\eta/2} + \frac{G^2}{4C} e^{-C\eta} + O\left(e^{-3C\eta/2}\right),\ \textrm{as} \ \eta \rightarrow \infty,
\label{eq:sakiadisA}
\end{equation}
where $B$, $Q$, $C$, and $G$ are constants arising from integration. We repeat the above expressions here to highlight a key difference and to motivate the technique used to examine the Sakiadis problem in this paper.  A necessary condition for an expansion to converge is that it does not introduce new singularities into the problem that lie in the physical domain.  In the Blasius problem, the constant $B$ takes on a negative value and thus~(\ref{eq:blasiusA}) has a singularity at the positive real value of $-B$. The expansion~(\ref{eq:sakiadisA}) for the Sakiadis problem does not have this issue, and it is thus possible that the expansion converges; this will be discussed in more detail shortly and revisited throughout the paper.

~\citet{Barlow:2017} used the asymptotic form~(\ref{eq:sakiadisA}) to propose an approximant to sum the divergent series~(\ref{eq:series})  for the Sakiadis problem as 
\begin{equation}
    \text{Sakiadis:  }f_{A,N}=C+\sum_{n=1}^N \mathcal{A}_n \left(e^{-C\eta/2}\right)^n.
    \label{eq:AA}
\end{equation}
In~(\ref{eq:AA}), the $\mathcal{A}_n$ coefficients are computed by solving an $N\times N$ system enforcing that the $N$-term Taylor expansion of~(\ref{eq:AA}) about $\eta=0$ is exactly the $N^\mathrm{th}$ truncation of~(\ref{eq:series}).  Unlike the recursive computation of $\tilde{b}_n$ in~(\ref{eq:Boyd}), all $\mathcal{A}_n$ coefficients change their value as $N$ changes, and thus~(\ref{eq:AA}) cannot be considered a formal power series solution to~(\ref{eq:Sakiadis}).  Nevertheless, $f_{A,N}$ converges to the numerical solution of $f$ over the physical domain as $N$ is increased.  Note that,  although not remarked on in~\citet{Barlow:2017}, each $\mathcal{A}_n$ coefficient converges to a specific value as $N$ is increased.
Additionally,~(\ref{eq:AA}) was used in~\cite{Barlow:2017} to compute $\kappa$, as well as the asymptotic constants in~(\ref{eq:sakiadisA}) ($C$ and $G$) to within 12 digits of accuracy (beyond previously reported numerical results) before hitting a round-off barrier. 

Note that, in using approximant~(\ref{eq:AA}),~\citet{Barlow:2017} did not utilize knowledge of the singularities that caused the original series~(\ref{eq:series}) to diverge. However, the convergence of~(\ref{eq:AA}) suggests that changing the gauge function from $\eta$ to $e^{-C\eta/2}$ fortuitously creates a mappping that removes the influence of the limiting singularities present in~(\ref{eq:series}); this is similar in behavior to the modified Euler transformation used in~(\ref{eq:Boyd}) for the Blasius problem. In the current work, we develop a formal Taylor series solution to the Sakiadis problem using the exponential gauge transformation mention above, with the goal of providing further explanation for the improved accuracy of~(\ref{eq:AA}). 

Interestingly, another classical problem in fluid mechanics that has a similar mathematical structure is a static air--liquid meniscus (with surface tension) formed when a flat wall is placed in an infinite horizontal pool. Both the meniscus and the Sakiadis problem have (a) the same asymptotic decaying exponential behavior at infinite distance from the wall and (b) a singularity structure that limits convergence of the standard power series solution, yet benefits from an exponential transformation. For the meniscus problem, the exact solution---given as distance from the wall as function of meniscus height---has long been known~\cite{Batchelor}.  Thus, the emphasis of this 2$^\mathrm{nd}$ analysis is not in providing a new solution form, but rather in exploiting the structural similarities with the Sakiadis problem to better understand the nature of exponential gauge transformations and their effect on convergence properties.

The paper is organized as follows. In section~\ref{sec:Sakiadis}, we review the series and approsimant solutions to the Sakiadis boundary layer problem, expanding upon the background given above.  In section~\ref{sec:SakiadisMapped}, an asymptotically motivated gauge function is used to map the Sakiadis ODE and B.C.s to a domain in which its series solution converges over the whole un-mapped physical domain.  In section~\ref{sec:predictions}, we provide a procedure for using the analytic solution (from section~\ref{sec:SakiadisMapped}) to compute the asymptotic constants $C$ and $G$, the wall shear parameter $\kappa$, and locations of the closest singularities (from $\eta$=0) of the Sakiadis function to arbitrary precision.  In section~\ref{sec:Meniscus}, the classical problem of a meniscus at a flat wall is reviewed and its solution is obtained in section~\ref{sec:divergent} as a divergent power series expansion about the wall location. An asymptotic expansion away from the wall and a convergent series solution motivated by this expansion are provided in section ~\ref{sec:MeniscusMapped}, following a similar technique to the Sakiadis problem in section~\ref{sec:SakiadisMapped}.   Remarks on the overall methodology applied to both problems are provided in section~\ref{sec:conclusions}. 

\section{The Sakiadis boundary layer}
\subsection{A divergent power series and a convergent approximant \label{sec:Sakiadis}}
Although the main results of this subsection are found in~\cite{Barlow:2017} (and references therein), here we elucidate elements of that work that were previously not discussed and are relevant to the work herein.  The Sakiadis boundary layer along a moving flat wall is obtained as a solution to the nonlinear ODE~(\ref{eq:Sakiadis})
where $f$ and $\eta$ are the similarity variables, related to the physical variables in the governing boundary layer equation~\citep{Sakiadis} through a similarity transform.  The general power series solution to the ODE~(\ref{eq:Sakiadis}a) was first developed by~\cite{Blasius}, and is expressed as
\begin{subequations}
\begin{equation}
    f(\eta) = \sum_{n = 0}^\infty a_n \eta^n,~~|\eta|<S
\end{equation} where\begin{equation}
    a_{n+3} = \frac{\ds -\sum_{j=0}^n (j+1)(j+2)a_{j+2}a_{n-j}}{2(n+1)(n+2)(n+3)},~~n\ge0. \label{SakTaylor0}
\end{equation}
\label{eq:SakSeries}
\end{subequations}

For the Sakiadis problem, $a_0=0$ and $a_1=1$ as prescribed by~(\ref{eq:Sakiadis}b), and $a_2=\kappa/2$ as defined in~(\ref{eq:kappa}). The ``wall shear'' parameter $\kappa$ is computed either numerically~\citep{Cortell} or analytically~\citep{Barlow:2017} in a self-consistent way such that the condition $f'(\infty)=0$ in~(\ref{eq:Sakiadis}b) is met. The value of $\kappa$ is given in~\cite{Barlow:2017} as
\begin{equation}
  \kappa=-0.443748313369\dots,
  \label{eq:kappaEst1}
\end{equation}
the precision of which is improved upon in section~\ref{sec:SakiadisMapped}. 

\begin{figure}[h!]
    \centering
    \subfloat{{\includegraphics[width=8cm]{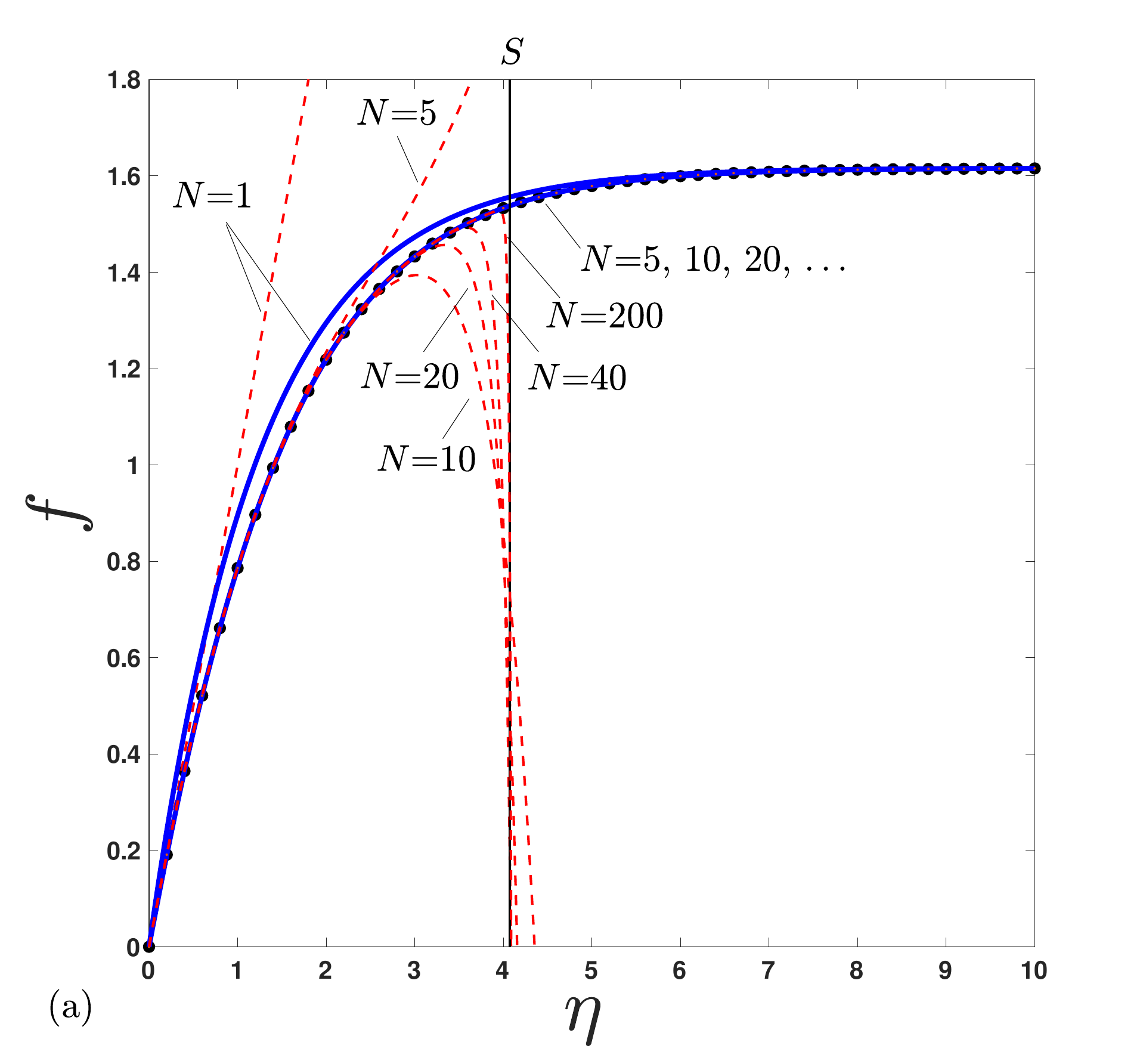} }}
    \subfloat{{\hspace{-0.2in}\includegraphics[width=8cm]{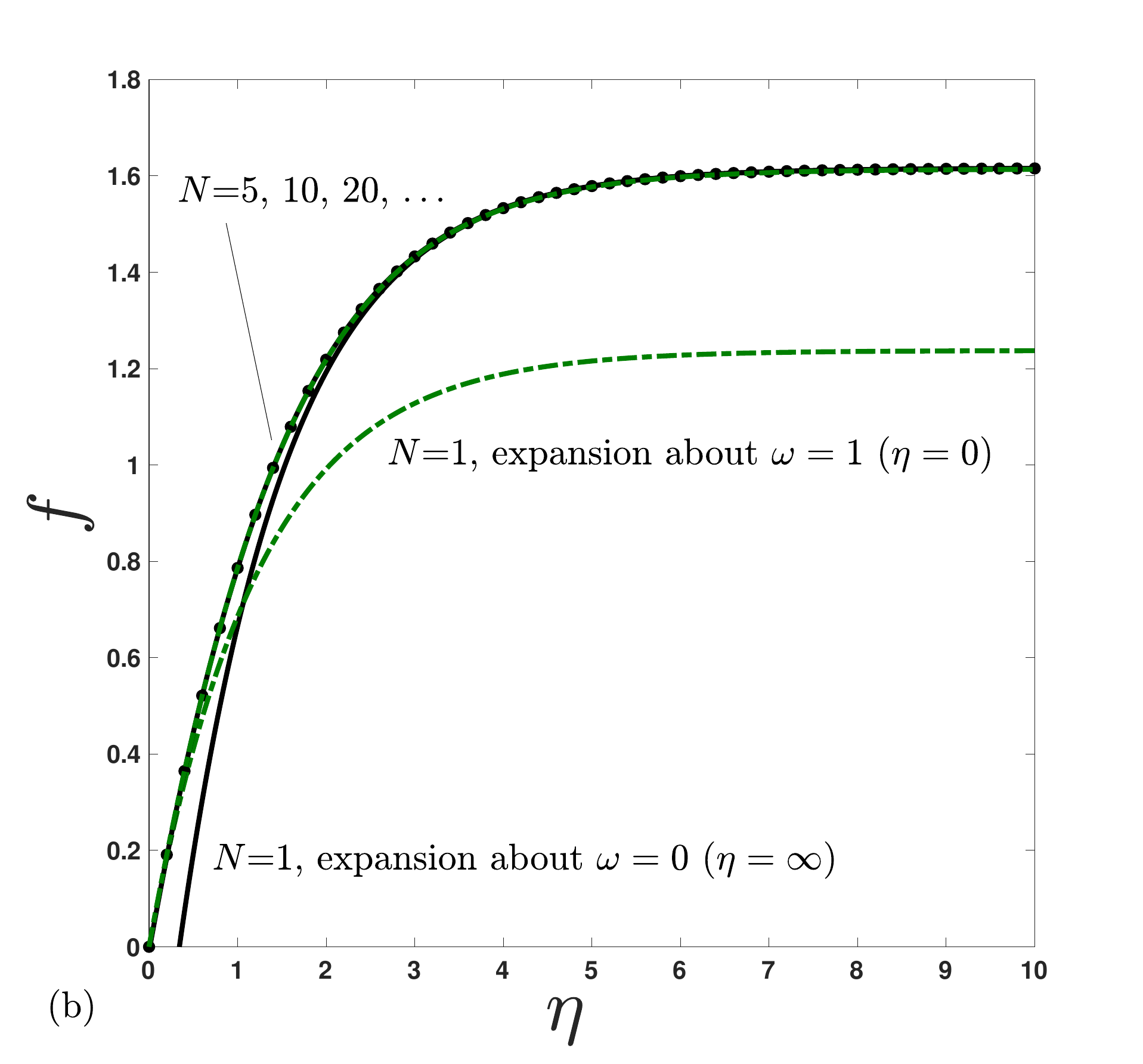} }}%
    \caption{Analytic solutions of~(\ref{eq:Sakiadis}) compared with the numerical solution (RK4 with $\Delta \eta=10^{-5}$) ($\bullet$'s).  (a) $N$-term truncations of series~(\ref{eq:SakSeries}) (\textcolor{red}{dashed curves}) and approximant~(\ref{SakApproxForm}) (\textcolor{blue}{solid curves}). A solid vertical line indicates the radius of convergence $S$ of~(\ref{eq:SakSeries}). (b) $N$-term truncations of series~(\ref{SakTransRec}) (solid curves) and series~(\ref{SakTransRec2}) (\textcolor{OliveGreen}{\--- - \---}) of section~\ref{sec:SakiadisMapped}.}%
    \label{fig:SakiadisUnMapped}%
\end{figure}

The series~(\ref{eq:SakSeries}) diverges within the physical domain, as indicated in figure~\ref{fig:SakiadisUnMapped}a (dashed lines), where it is compared to the numerical solution ($\bullet$'s, RK4 with $\Delta \eta=10^{-5}$).   Using the modified Pad\'e approximants given in~\cite{Barlow:2017}, an estimate of the locations of the  convergence-limiting singularities in the complex plane are found\footnote{In~\cite{Barlow:2017}, only the singularity at $-1.2114 + 3.8878i$ is reported.  Revisiting the analysis, we find two things: (1) The expected conjugate $-1.2114 - 3.8878i$ is also found in the roots of the denominator of the modified Pad\'e and (2) The two additional digits of $\eta_s$ reported in~\cite{Barlow:2017} (but not here) are found to converge to different values as more terms are taken; a more precise value is given here in section~\ref{sec:predictions}.} to be at $\eta_{s\pm} \approx -1.2114 \pm 3.8878i $, indicating a radius of convergence of $S=|\eta_{s\pm}| \approx 4.0722$, shown in figure~\ref{fig:SakiadisUnMapped}a as a vertical line. 

The series~(\ref{eq:SakSeries}) is analytically continued beyond this radius of convergence in~\cite{Barlow:2017} in an approximate way by constructing an approximant based on the $\eta\to\infty$ expansion given by~(\ref{eq:sakiadisA}). As mentioned in section~\ref{sec:intro}, a particular sequence of approximants having the form of the asymptotic behavior~(\ref{eq:sakiadisA}) are examined in~\cite{Barlow:2017} as
\begin{equation}
    f_{A, N}(\eta) = C+\sum_{k=1}^N \mathcal{A}_k \left(e^{-C\eta /2}\right)^k \label{SakApproxForm}
\end{equation}
\begin{equation}
\left[ \begin{array}{cccc}

1 &&& 1  \\ 1 &&& 2 \end{array} \right]\left[ \begin{array}{c}

\mathcal{A}_1\\ \mathcal{A}_2 \end{array} \right]=\left[ \begin{array}{c}

0!~a_0-C\\ 1!~(-2/C)~a_1  \end{array} \right],
\label{matrix2}
\end{equation}
and for $N$=3 we have 
\begin{equation}
\left[ \begin{array}{ccccccc}

1 &&& 1 &&& 1  \\ 1 &&& 2 &&& 3 \\ 1 &&& 4 &&& 9 \end{array} \right]\left[ \begin{array}{c}

\mathcal{A}_1\\ \mathcal{A}_2 \\ \mathcal{A}_3 \end{array} \right]=\left[ \begin{array}{c}

0!~a_0-C\\ 1!~(-2/C)~a_1 \\ 2!~ (-2/C)^2 ~a_2 \end{array} \right], 
\label{matrix3}
\end{equation}
 and so on.  Note that, after inverting the matrices above,  $\mathcal{A}_1$ and $\mathcal{A}_2$ attain different values in~(\ref{matrix2}) than in~(\ref{matrix3}).  The relevance of this issue is discussed in what follows.   Also, although the Vandermonde matrix has an explicit inversion formula, high precision arithmetic (i.e., beyond double) must be used for large $N$ to avoid round-off error. 

In principle, if $\kappa$ and $C$ are known,~(\ref{SakApproxForm}) provides a complete approximate solution. A slight modification can also be used to estimate $\kappa$ and $C$ as well, wherein the $\mathcal{A}_k$ coefficients are chosen to match the Taylor coefficients to order $N$, but $\kappa$ and $C$ are chosen to make $\mathcal{A}_N=\mathcal{A}_{N-1}=0$ (effectively switching $\mathcal{A}_N$ and $\mathcal{A}_{N-1}$ with $\kappa$ and $C$ as unknowns of the $N\times N$ system). This is precisely what is done in~\cite{Barlow:2017} to compute the value of $\kappa$ given by~(\ref{eq:kappaEst1}) and to provide a $C$ value of 
\begin{equation}
    C=1.61612544681\dots;
    \label{CEst1}
\end{equation}
the precision of $C$ is improved upon in section~\ref{sec:SakiadisMapped}. After using the $\kappa$ and $C$ obtained via the above process, approximant~(\ref{SakApproxForm}) converges uniformly over the domain as $N$ is increased, as shown in figure~\ref{fig:SakiadisUnMapped}a (solid curves).  Curves for $f'$ and $f''$, found analytically from~(\ref{SakApproxForm}), are also shown to uniformly converge in~\cite{Barlow:2017}. To revisit the discussion just above, it is found that, as the approximant~(\ref{SakApproxForm}) itself converges and the values of $A_k$ change with increasing $N$, $A_1$ and $A_2$ converge respectively to values
\begin{equation}
   A_1=G=-2.1313459241\dots
   \label{Gest1}
 \end{equation}
 and $A_2=G^2/(4C)$, matching the $\eta\to\infty$ expansion~(\ref{eq:sakiadisA}) obtained from dominant balance.  The other $A_k$ coefficients also converge, albeit slower as $k$ increases, as the portion of the approximant that is used to match the behavior at $\eta=0$ is pushed to higher-order terms.  

Although the approximant~(\ref{SakApproxForm}) is a \textit{convergent} series, its coefficients depend on truncation $N$ and so it cannot be considered a formal \textit{Taylor} series. Thus, we cannot apply Taylor's theorem to~(\ref{SakApproxForm}) to relate convergence and singularity location in the context of switching between a gauge function of $\eta$ in~(\ref{eq:SakSeries}) to a gauge function of $e^{-C\eta/2}$ in~(\ref{SakApproxForm}).  In the next section, we remove this coefficient dependence on truncation while retaining the asymptotically motivated gauge function of $e^{-C\eta/2}$;  in doing so, we provide an explanation for why the approximant~(\ref{SakApproxForm}) converges without explicitly incorporating singular behavior.  

\subsection{Variable transform and convergent power series solution \label{sec:SakiadisMapped}}
Motivated by the asymptotic expansion~(\ref{eq:sakiadisA}) and encouraged by the convergence of~(\ref{SakApproxForm}), we define the variable transformations
\begin{align}
    \omega(\eta) &= \exp\left(-\frac{C}{2} \eta\right) \label{SakTransDefs}
    \end{align}
    \begin{align}
    f(\eta) \equiv  g(\omega(\eta)).
    \label{composition}
\end{align}

Substituting~(\ref{SakTransDefs}) and~(\ref{composition}) into~(\ref{sDE}), applying the chain rule, and rearranging terms, the transformed ODE becomes
\begin{equation}
    \frac{g}{C}\left(\omega\ddot{g} + \dot{g}\right) = \omega^2 \dddot{g} + 3\omega \ddot{g} + \dot{g}, \label{SakTransformedODE}
\end{equation}
where $\dot{g}$ denotes a derivative with respect to $\omega$.  The boundary conditions~(\ref{sBC}) become
\begin{equation}
    g(1)=0,~~\dot{g}(1)=\frac{-2}{C},~~\ddot{g}(1)=\frac{2}{C}+\frac{4\kappa}{C^2}
    \label{eq:transcond}
\end{equation}
where the third condition in~(\ref{sBC}) here leads to $0=0$ and so the transformation of condition~(\ref{eq:kappa}) is written instead for $\ddot{g}(1)$. 

In the usual way, we assume a solution to~(\ref{SakTransformedODE}) of the form \begin{equation}
    g(\omega) = \sum_{n=0}^\infty \tilde{a}_n \omega^n,
    \label{eq:omega}
\end{equation} 
which is readily differentiated term-by-term to compute $\dot{g}$, $\ddot{g}$, and $\dddot{g}$.  After employing Cauchy's product rule~\citep{Churchill} to handle the nonlinear term on the left-hand side of~(\ref{SakTransformedODE}), the ODE becomes 

\begin{multline}
    \frac{\omega}{C}\sum_{n=0}^\infty \left[\sum_{k=0}^n (k+1)(k+2)\tilde{a}_{k+2}\tilde{a}_{n-k}\right]\omega^n + \frac{1}{C}\sum_{n=0}^\infty \left[\sum_{k=0}^n (k+1)\tilde{a}_{k+1}\tilde{a}_{n-k}\right]\omega^n \\ = \omega^2 \sum_{n=0}^\infty (n+1)(n+2)(n+3)\tilde{a}_{n+3}\omega^n + 3\omega \sum_{n=0}^\infty (n+1)(n+2)\tilde{a}_{n+2}\omega^n \\ + \sum_{n=0}^\infty (n+1)\tilde{a}_{n+1}\omega^n. \label{SakTransformedExpandedForm}
\end{multline}

Equating constant terms on both sides of~(\ref{SakTransformedExpandedForm}) leads to 
\begin{subequations}
\begin{equation}
   \tilde{a}_0=C.
   \label{eq:Cdef}
\end{equation}
Equating $\omega^1$ terms on both sides of~(\ref{SakTransformedExpandedForm}) leads to $\tilde{a}_1=\tilde{a}_1$ and to match the notation of the asymptotic expansion~(\ref{eq:sakiadisA}), we write
\begin{equation}
   \tilde{a}_1=G.
\end{equation}
Equating $\omega^n$ terms for $n\geq 1$ on both sides of~(\ref{SakTransformedExpandedForm}) leads to \[\sum_{k=0}^{n-1}(k+1)(k+2)\tilde{a}_{k+1}\tilde{a}_{n-k-1}+\sum_{k=0}^n(k+1)\tilde{a}_{k+1}\tilde{a}_{n-k}=C(n+1)^3\tilde{a}_{n+1},\]
which may be simplified (by appropriate shifts to the $k$ indices) as 
\[\sum_{k=1}^{n+1}k^2 \tilde{a}_k \tilde{a}_{n-k+1}=C(n+1)^3\tilde{a}_{n+1}.\]
Solving the above expression for $\tilde{a}_{n+1}$ and using~(\ref{eq:Cdef}) leads to the recursion
\begin{equation}
    \tilde{a}_{n+1} = \frac{1}{C n (n+1)^2}\sum_{k=1}^{n}k^2 \tilde{a}_k \tilde{a}_{n-k+1}. \ n \geq 1. \label{SakTransformedRecurrence}
\end{equation}
 Note that for $n=1$,~(\ref{SakTransformedRecurrence}) leads to 
\[
\tilde{a}_2 = \ds \frac{G^2}{4C},
\]
which matches the asymptotic expansion~(\ref{eq:sakiadisA}).  
The transformed solution to~(\ref{eq:Sakiadis}), written in terms of the original variables is then
\begin{equation}
    f(\eta) = \sum_{n=0}^\infty \tilde{a}_n \left(e^{-C\eta/2}\right)^n. 
\end{equation}
\label{SakTransRec}
\end{subequations}

As we can see,~(\ref{SakTransRec}) is a Taylor series solution to~(\ref{eq:Sakiadis}) in terms of the gauge function $e^{-C\eta /2}$. Although the implementation of~(\ref{SakTransRec}) requires knowledge of $C$ and $G$, we will show in section~\ref{sec:predictions} that these may be computed (to any desired precision) directly from~(\ref{SakTransRec}) via the conditions~(\ref{eq:transcond}).  Interestingly,~(\ref{SakTransRec}) appears to be a recursive way to compute the asymptotic series~(\ref{eq:sakiadisA}), originally obtained in~\cite{Barlow:2017} via the method of dominant balance.
\begin{figure}[h!]%
    \centering
    \subfloat{{\includegraphics[width=8cm]{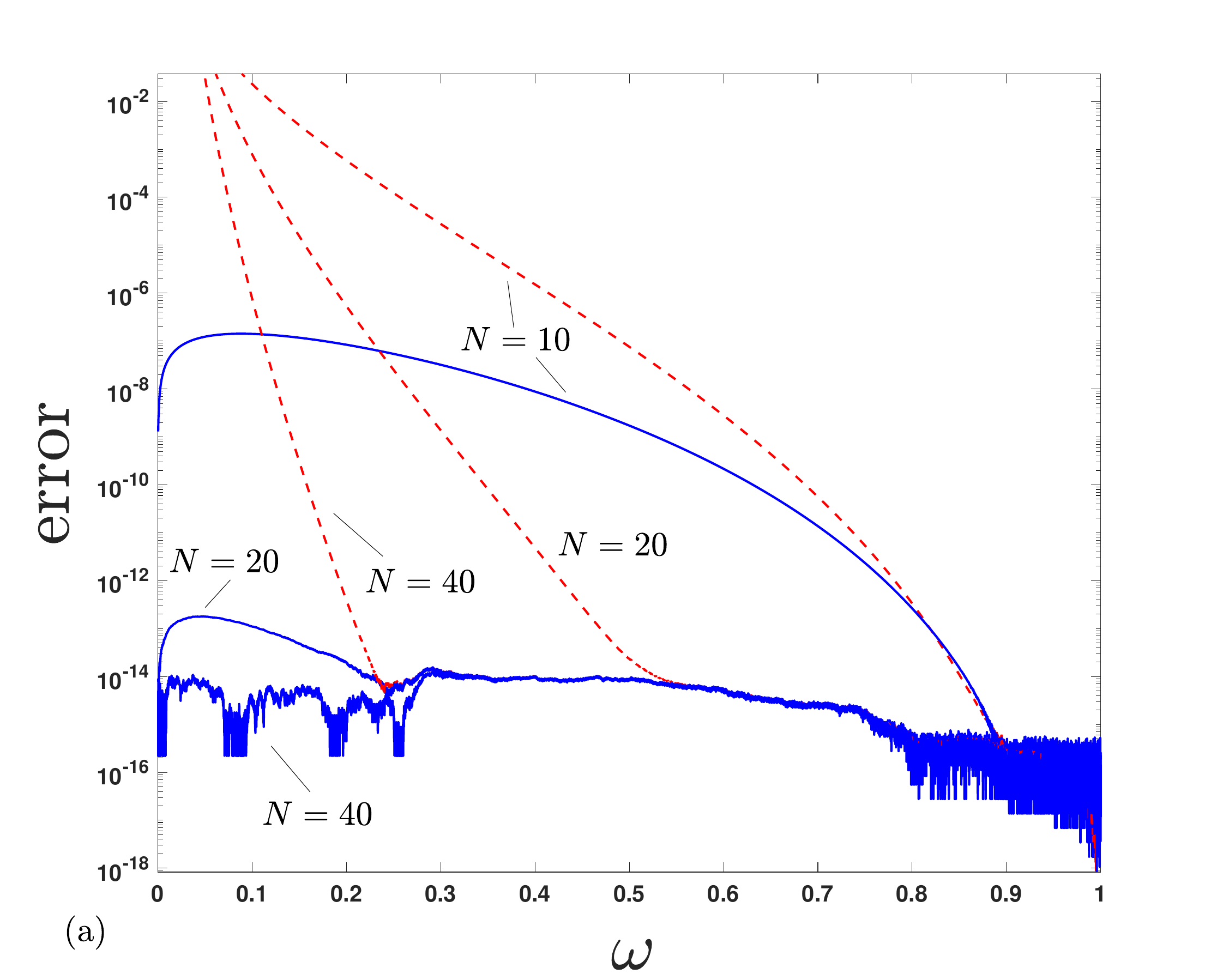} }}
    \subfloat{\hspace{-0.2in}{\includegraphics[width=8cm]{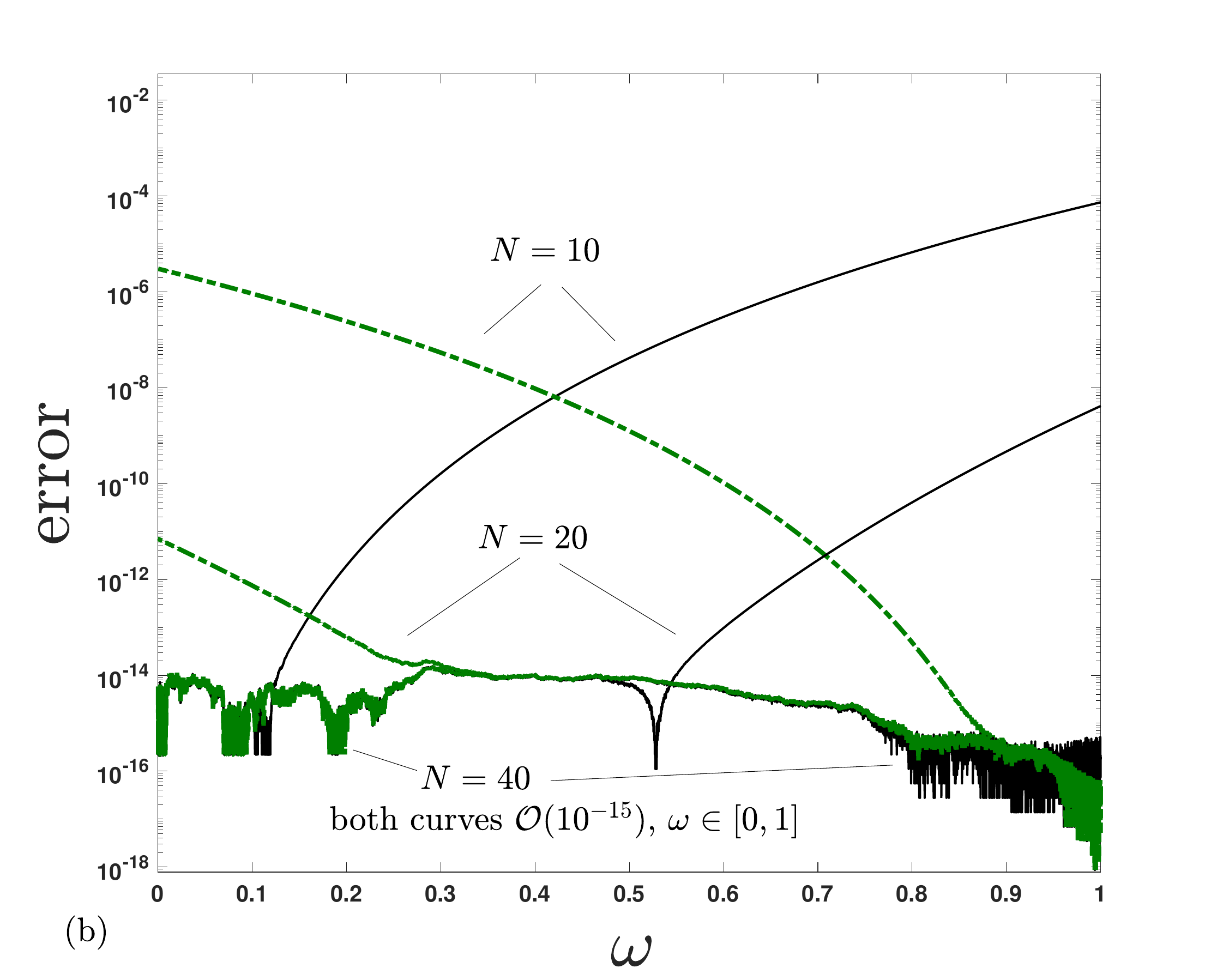} }}%
    \caption{Absolute error between analytical solutions and the numerical solution of~(\ref{eq:Sakiadis}) shown in figure~\ref{fig:SakiadisUnMapped}, here plotted versus transformed variable $\omega$ to encompass the full physical domain $\omega\in[0, 1]$. (a) $N$-term truncations of series~(\ref{eq:SakSeries}) (\textcolor{red}{dashed curves}) and approximant~(\ref{SakApproxForm}) (\textcolor{blue}{solid curves}).  (b) $N$-term truncations of series~(\ref{SakTransRec}) (solid curves) and series~(\ref{SakTransRec2}) (\textcolor{OliveGreen}{\--- - \---}) of section~\ref{sec:SakiadisMapped}.}%
    \label{fig:SakiadisErrorMapped}%
\end{figure}

Using values of $C$ and $G$ determined in section~\ref{sec:predictions}, convergence of~(\ref{SakTransRec}) is shown as solid lines in figure~\ref{fig:SakiadisUnMapped}b.  An error plot is shown in figure~\ref{fig:SakiadisErrorMapped}b, indicating convergence to within machine precision over the full domain (shown versus $\omega\in[0,1]$ for clearer representation of the limit $\eta\to\infty$). For comparison, error from the divergent Sakiadis series~(\ref{eq:SakSeries}) and approximant~(\ref{SakApproxForm}) are shown in figure~\ref{fig:SakiadisErrorMapped}a over the same domain. Although not shown here, error for $f'$ and $f''$, which may be analytically obtained from~(\ref{SakTransRec}), follow the same error trends as in figure~\ref{fig:SakiadisErrorMapped}b.

\begin{figure} [h!]
    \centering
    \subfloat{{\includegraphics[width=7.2cm]{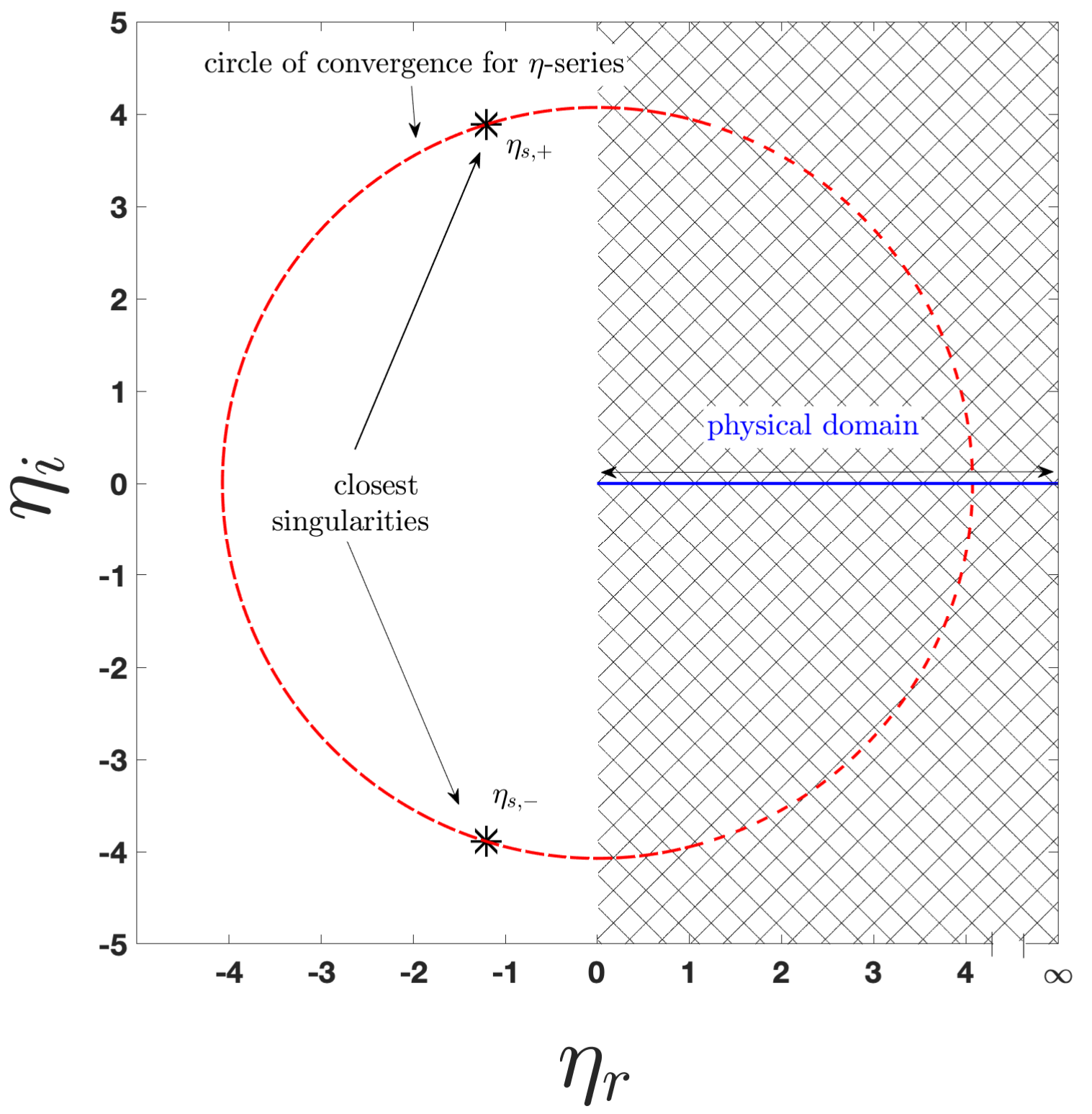} }}%
    \subfloat{{\includegraphics[width=7.6cm]{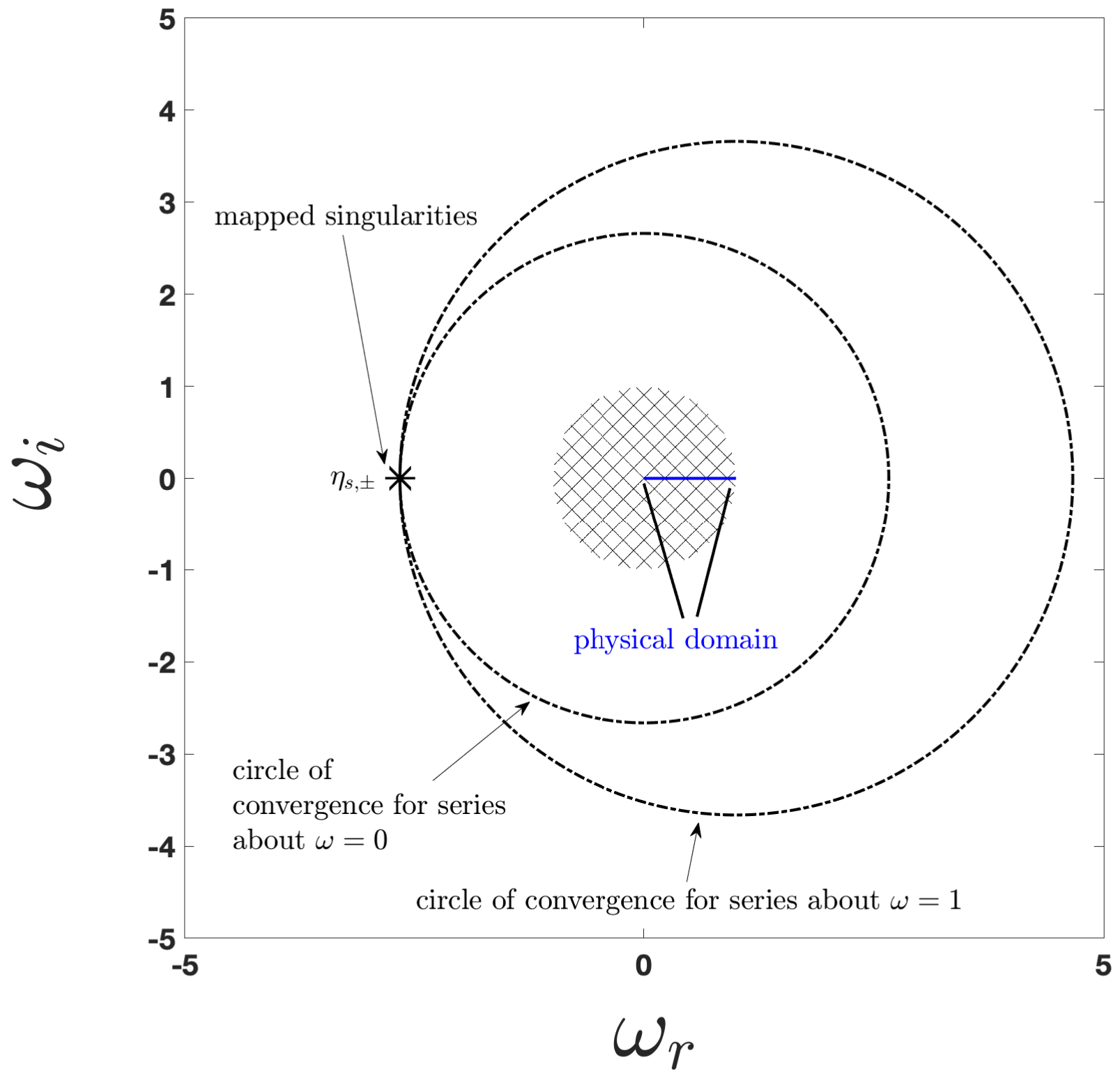} }}%
    \caption{The effect of the gauge transformation~(\ref{SakTransDefs}) on $\eta$-singularities ($\ast$) of the Sakiadis function, showing their placement in the 2$^\mathrm{nd}$ and 3$^\mathrm{rd}$ quadrants of the complex $\eta$ plane (left) and their image on the negative real $\omega$ line (right). The interior of the \textcolor{red}{dashed circle} shown in the $\eta$ plane delineates the region of convergence of series~(\ref{eq:SakSeries}).  The interiors of the smaller and larger dashed circles shown in the $\omega$ plane respectively delineate the regions of convergence of~(\ref{SakTransRec}) and~(\ref{SakTransRec2}).}%
    \label{fig:SakiadisMapped}%
\end{figure}

We now provide an explanation of why~(\ref{SakTransRec}) (and by extension~(\ref{SakApproxForm})) converge to the numerical solution of~(\ref{eq:Sakiadis}). As alluded to in section~\ref{sec:intro}, the singularities that cause the original series~(\ref{eq:SakSeries}) to diverge are mapped via the gauge function~(\ref{SakTransDefs}) such that their influence lies beyond the physical domain; this is shown in figure~\ref{fig:SakiadisMapped} where the complex $\eta$ and $\omega$ planes are compared. Here, we track the movement of the two  singularities $\eta_{s,\pm}$ (shown by $\ast$'s in the figure) closest to $\eta=0$, whose locations are predicted by the Pad\'e analysis of~\cite{Barlow:2017} (see section~\ref{sec:Sakiadis}) and later confirmed in section~\ref{sec:predictions}. In the $\eta$ plane of figure~\ref{fig:SakiadisMapped}, the circle of convergence for~(\ref{eq:SakSeries}) is drawn based on these two singularities.  Note that this circle intersects the positive real line at the same location as the radius of convergence of series~(\ref{eq:SakSeries}) drawn in figure~\ref{fig:SakiadisUnMapped}a, as is expected from Taylor's theorem, which only guarantees convergence of Taylor series within such circles~\citep{Watson}.   In the $\omega$ plane of figure~\ref{fig:SakiadisMapped}, the circle of convergence for~(\ref{SakTransRec}) is drawn, centered at $\omega=0$ and with mapped radius $|e^{-\left(C\eta_{s,\pm}\right) /2}|$; note that this circle now extends beyond the physical domain itself\footnote{To paraphrase Buzz Lightyear:  ``To $\eta=\infty$ and beyond''}, which explains why~(\ref{SakTransRec}) converges over the entire physical domain.  

\begin{figure}%
    \centering
    \subfloat{{\includegraphics[width=9cm]{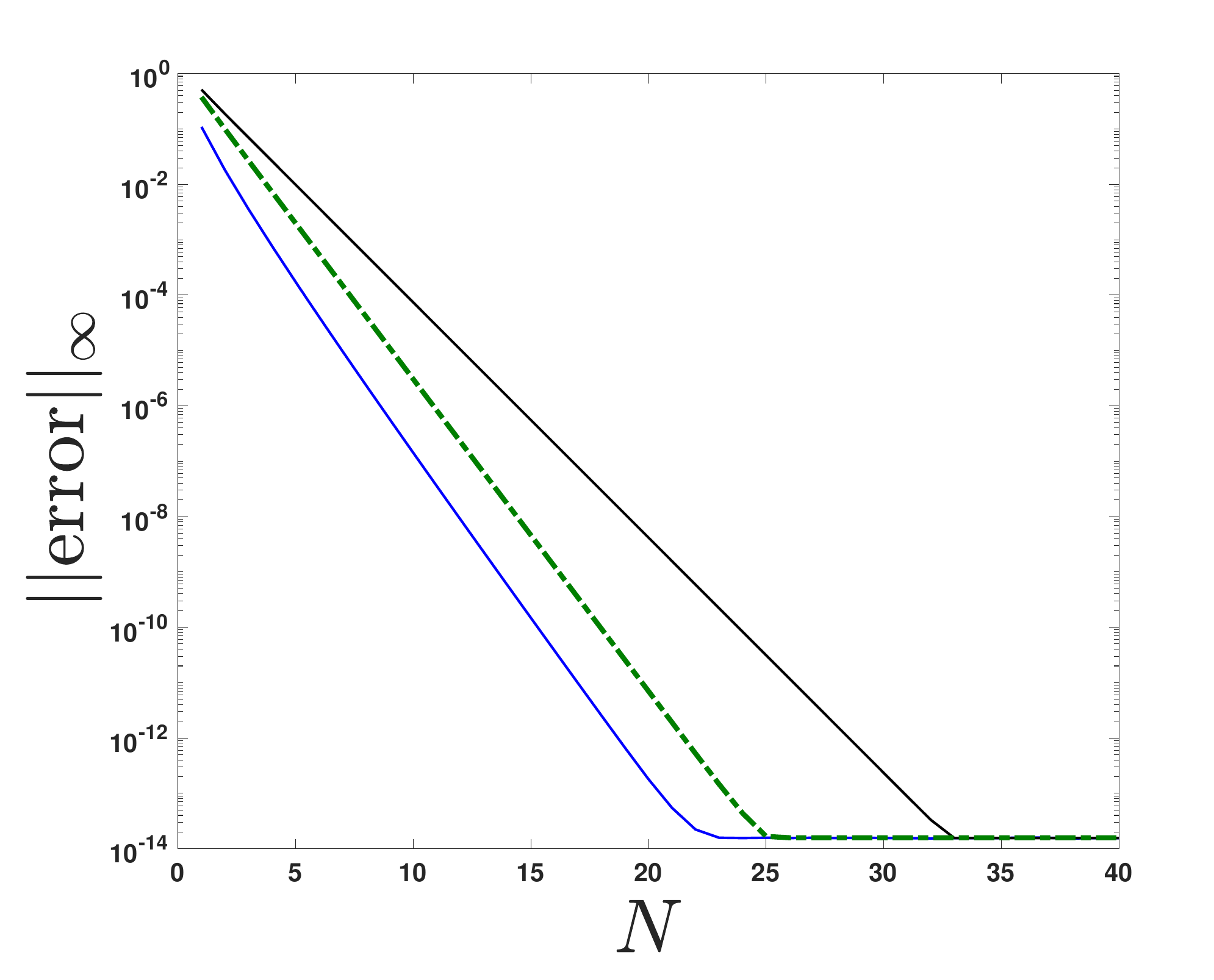} }}%
    \caption{The infinity norm of the difference between convergent analytical solutions and the numerical solution to~(\ref{eq:Sakiadis}), taken over the domain shown in figure~\ref{fig:SakiadisErrorMapped} and plotted versus series truncation $N$.  The legend is the same as in previous figures, showing~(\ref{SakApproxForm}) series (lowermost \textcolor{blue}{solid curve}), series~(\ref{SakTransRec2}) (middle \textcolor{OliveGreen}{\--- - \--- curve}), and series~(\ref{SakTransRec}) (uppermost solid curve). The trends shown here  are the same for the norm of the error of $f'$ and $f''$ using~(\ref{SakApproxForm}),~(\ref{SakTransRec2}), and~(\ref{SakTransRec}).}%
    \label{fig:SakiadisNorm}%
\end{figure}

On the mapped $\omega$-plane shown in figure~\ref{fig:SakiadisMapped}, note that, in addition to drawing the circle of convergence corresponding to expansion~(\ref{SakTransRec}), a larger circle corresponding to the expansion about $\omega=1$ is also drawn, which is larger because the mapped singularities are farther away.  This expansion point corresponds to $\eta=0$ in the original domain.  Using the same procedures employed above to obtain the series about $\omega=0$, the series about $\omega=1$ is defined as
\begin{equation}
    g(\omega) = \sum_{n=0}^\infty \hat{a}_n (\omega-1)^n
    \label{eq:g1}
\end{equation}
where the coefficients are given as
\begin{subequations}
\begin{align}
\nonumber
    \hat{a}_{n+3} = &-\frac{2n+3}{n+3}\hat{a}_{n+2}-\frac{(n+1)^2}{(n+2)(n+3)}\hat{a}_{n+1}\\
    \nonumber&+\frac{\displaystyle\sum_{k=0}^{n-2}(k+1)(k+2)\hat{a}_{k+2}\hat{a}_{n-k-1} + \sum_{k=0}^{n-1} (k+1)\left[(k+2)\hat{a}_{k+2}+\hat{a}_{k+1}\right]\hat{a}_{n-k}}{C(n+1)(n+2)(n+3)}, \ n \geq 0\\
    &\hat{a}_0=0,~~\hat{a}_1=-2/C,~~\hat{a}_2=2\kappa/C^2+1/C
    \label{SakTransRec2}
\end{align}
and, writing~(\ref{eq:g1}) in terms of the original variables, we have
\begin{equation}
    f(\eta) = \sum_{n=0}^\infty \hat{a}_n\left(e^{-C\eta/2}-1\right)^n,
\end{equation}
\label{SakTransRec2}
\end{subequations}
which may be considered a series solution to~(\ref{eq:Sakiadis}) written in terms of the gauge function $\left(e^{-C\eta/2}-1\right)$.  In figure~\ref{fig:SakiadisErrorMapped}b, the error associated with~(\ref{SakTransRec2}) (dashed line) is shown to reduce to machine precision as additional terms are used in the series. Although it is difficult to identify an advantage in  using~(\ref{SakTransRec2}) over~(\ref{SakTransRec}), as each naturally perform better near their expansion point as they converge, a ``best choice'' can be chosen by examining the rapidity of convergence.  This can be determined by considering the infinity norm of the error (over $\omega\in[0,1]$) versus truncation $N$, shown in figure~\ref{fig:SakiadisNorm}. In the figure, one can see that, for a fixed number of terms $N$, the expansion about $\omega=1$ (given by~(\ref{SakTransRec2}))  is more accurate than that about $\omega=0$ (given by~(\ref{SakTransRec})).  For comparison, the approximant~(\ref{SakApproxForm}) is also shown in figure~\ref{fig:SakiadisNorm} and outperforms both~(\ref{SakTransRec}) and~(\ref{SakTransRec2}). The cost in using~(\ref{SakApproxForm}) is computational, since either inverting an increasingly ill-conditioned Vandermonde matrix or using a Vandermonde inversion algorithm requires precision well-beyond double to avoid round-off error. 

\subsection{Computation of $C$, $G$, $\kappa$, and singularities of the Sakiadis function \label{sec:predictions}}

Previous estimates for $C$, $G$, and $\kappa$ have been found by quadrature, shooting, and approximant~(\ref{SakApproxForm})~\citep{Eftekhari,Cortell,Barlow:2017}. Here, we improve upon those estimates and provide formulae for obtaining them to any desired precision, by considering the application of conditions~(\ref{eq:transcond}) to the convergent solution~(\ref{SakTransRec}), i.e.,
\begin{align}
    &g(1)=\sum_{n=0}^\infty \tilde{a}_n=0 \label{eq1}\\
    &\dot{g}(1)=\sum_{n=1}^\infty n\tilde{a}_n=\frac{-2}{C} \label{eq2}\\
    &\ddot{g}(1)=\sum_{n=2}^\infty n(n-1)\tilde{a}_n=\frac{2}{C}+\frac{4\kappa}{C^2}.
    \label{eq3}
\end{align}
Noting that~(\ref{eq1}) and~(\ref{eq2}) are functions of $C$ and $G$ only (see~(\ref{SakTransRec})), we have 2 equations and 2 unknowns and one may employ any desired solver (here we use Newton's method). After finding $C$ and $G$, one may then use~(\ref{eq3}) to explicitly compute $\kappa$. In preparation for Newton's method, we rewrite~(\ref{eq1}) and~(\ref{eq2}) as 
\begin{align}
    \phi_0(C, G) &= \sum_{n=0}^\infty \tilde{a}_n=0 \\ \phi_1(C, G) &= \frac{2}{C} + \sum_{n=1}^\infty n\tilde{a}_n=0,
\end{align}
such that the $\nu^\mathrm{th}$ Newton iterate becomes
\begin{subequations}
\label{eq:Newton}
\begin{equation}
    \left[ \begin{array}{c}

C_{\nu+1}  \\ G_{\nu+1} \end{array} \right]= \left[ \begin{array}{c}

C_{\nu}  \\ G_{\nu} \end{array} \right]-\left[ \begin{array}{cc}

\left(\frac{\partial\phi_0}{\partial C}\right)_\nu &  \left(\frac{\partial\phi_0}{\partial G}\right)_\nu\\ \left(\frac{\partial\phi_1}{\partial C}\right)_\nu &  \left(\frac{\partial\phi_1}{\partial G}\right)_\nu \end{array} \right]^{-1}\left[ \begin{array}{c}

\phi_{0,\nu}\\ \phi_{1,\nu} \end{array} \right].
\label{eq:NewtonIterate}
\end{equation}
The partial derivatives in~(\ref{eq:NewtonIterate}) may be computed compactly by recognizing the pattern in the series~(\ref{SakTransRec}) as
\[g=C+G\omega+ \frac{1}{4}\frac{G^2}{C}\omega^2+ \frac{5}{72}\frac{G^3}{C^2}\omega^3+\dots+ \underbrace{a'_nG^nC^{1-n} }_{\tilde{a}_n} \omega^n +\dots\]
where $a'_n$ is $\tilde{a}_n$ evaluated at $C=G=1$ (i.e., the  coefficient with $C$ and $G$ removed)\footnote{Although here we claim this form of $C$ and $G$ dependence on $\tilde{a}_n$ by inspection, this can be verified by making the substitutions $\omega=G\tilde{\omega}/C$ and $g=C\tilde{g}$ into~(\ref{SakTransformedODE}) and noting the $C$ and $G$ dependence vanishes.}.  Using the dependence shown above, the partial derivatives needed for~(\ref{eq:NewtonIterate}) are found\footnote{In cases where the parameter dependence is more complicated, this can always be done recursively by directly partially differentiating the recurrence relation.} by first partially differentiating each coefficient with respect to $G$ or $C$, noting that $a_n'$ is a constant, and multiplying and dividing by $G$ or $C$ to make the expression in terms of $\tilde{a}_n$. Thus we obtain
\begin{align}
\nonumber
    \frac{\partial \phi_0}{\partial C} &= \frac{1}{C}\sum_{n=0}^\infty  (1-n)\tilde{a}_n && \frac{\partial \phi_0}{\partial G} = \frac{1}{G}\sum_{n=0}^\infty n\tilde{a}_n \\ \frac{\partial \phi_1}{\partial C} &= \frac{-2}{C^2}+\frac{1}{C} \sum_{n=1}^\infty  n (1-n)\tilde{a}_n && \frac{\partial \phi_1}{\partial G} = \frac{1}{G}\sum_{n=1}^\infty  n^2\tilde{a}_n.
    \label{eq:newtonseries}
\end{align}
\end{subequations}

Using the method outlined above, the following improved estimates are obtained beyond double precision using only a few Newton iterations (if previous estimates are used as the initial guess) in~(\ref{eq:Newton}) and a truncation of 50 terms in the series of~(\ref{eq:newtonseries}): 

\[C=1.616125446804603717\dots\]

\[G=-2.1313459240475714821 \dots\]

Using the above values of $C$ and $G$, we add the left and right side of (\ref{eq2}) and (\ref{eq3}), extract $\tilde{A}_1 = G$ from (\ref{eq2}), and compute $\kappa$ as 
\begin{equation}
    \kappa=\frac{C^2}{4}\left[G+\sum_{n=2}^\infty n^2\tilde{a}_n\right],
    \label{getkappa}
\end{equation}
where, taking 50 terms in~(\ref{getkappa}), we obtain  
\[\kappa=-0.443748313368861 \dots\]

\begin{figure}[h!]
    \centering
    \subfloat{{\includegraphics[width=9cm]{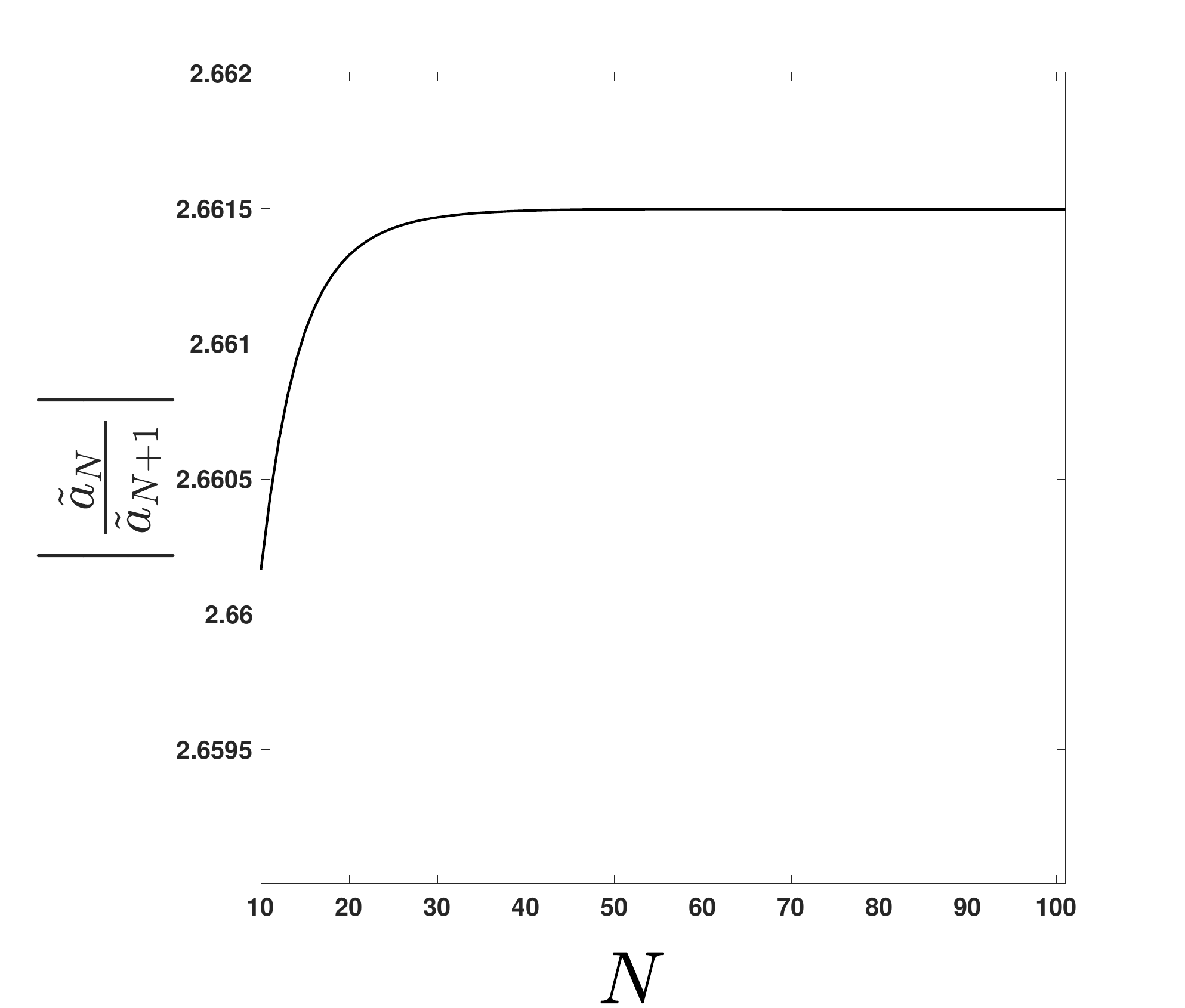} }}%
    \caption{The limiting behavior of the ratio test for series~(\ref{SakTransRec}) versus truncation $N$}%
    \label{fig:SakiadisROC}%
\end{figure}

Using the values of $C$ and $G$ given above, we may now explore the convergence of~(\ref{SakTransRec})  with the aim of deducing the precise singularity locations sketched in figure~\ref{fig:SakiadisMapped}.  As shown in figure~\ref{fig:SakiadisROC}, the radius of convergence of~(\ref{SakTransRec}) can be deduced from the ratio test, and is approximately\[ \lim_{N\to\infty}\left|\frac{\tilde{a}_N}{\tilde{a}_{N+1}}\right| = 2.66149513\dots\]
Relating the above to the original variables of the Sakiadis problem via~(\ref{SakTransDefs}), we have 
\begin{align}
\nonumber
    \lim_{N\to\infty}\left|\frac{\tilde{a}_N}{\tilde{a}_{N+1}}\right|&=\left|e^{-C\left(\eta_{s,r}\pm  i \eta_{s,i}\right)/2}\right|\\
    &=e^{-C\eta_{s,r}/2}
    \label{eq:ratio}
\end{align}
where $\eta_{s,r}$ and $\eta_{s,i}$ are the respective real and imaginary parts of the conjugate singularity pair closest to $\eta=0$ in the original Sakiadis function. While~(\ref{eq:ratio}) may be used to solve directly for $\eta_{s,r}$, the imaginary part may be deduced by recognizing that~(\ref{SakTransRec}) appears (for all terms investigated) to be an alternating series and thus the closest $\omega$-singularity is on the negative real line\footnote{This follows from a variable substitution $\tilde{\omega}=-\omega$ in~(\ref{eq:omega}), application of Pringsheim's theorem~\citep{Flajolet} to the resulting series, and then mapping back to $\omega$.} (indicated in figure~\ref{fig:SakiadisMapped}b) such that  
\begin{equation}
    \arg\left[e^{-C\left(\eta_{s,r}\pm  i \eta_{s,i}\right)/2}\right]=\mp C\eta_{s,i}/2=\mp(2n-1)\pi,~~n=1,2,3\dots.
    \label{eq:arg}
\end{equation}
where, choosing $n=1$ in~(\ref{eq:arg}) leads to $\eta_{s,i}$ values consistent with both the Pad\'e analysis in~\cite{Barlow:2017} and the radius of convergence of~(\ref{eq:SakSeries}) indicated in figure~\ref{fig:SakiadisUnMapped} (when combined with $\eta_{s,r}$ above).
Hence we conjecture that $n$ equals 1 in~(\ref{eq:arg}) and, consequently, that the closest singularities to $\eta=0$ in the Sakiadis problem are given, through use of~(\ref{eq:ratio}) and~(\ref{eq:arg}), by
\begin{align}
\nonumber
    \eta_{s,\pm}&=\frac{-2}{C}\lim_{N\to\infty}\ln\left|\frac{\tilde{a}_N}{\tilde{a}_{N+1}}\right|\pm i\frac{2\pi}{C}\\&\approx-1.211401\pm3.887808i.
    \label{eq:SakSing}
\end{align}

It is worth mentioning that, although we are able to use the convergent series solution~(\ref{SakTransRec}) to deduce the locations of the convergence-limiting singularities of the divergent series~(\ref{eq:SakSeries}), knowledge of the singularity~(\ref{eq:SakSing}) does not enter into the formulation of our solution at any point, in contrast with the modified Euler summation~(\ref{eq:Boyd}) used for the Blasius problem. Consequently (and fortunately), the convergence of~(\ref{SakTransRec}) does not rely on retaining ``enough'' digits of~(\ref{eq:SakSing}).  

In section~\ref{sec:Meniscus}, an exponential transformation similar to~(\ref{SakTransDefs}) is used to obtain a convergent series from a divergent one. However, in contrast with the Sakiadis problem, the meniscus problem of section~\ref{sec:Meniscus} has an exact solution that allows us to anticipate the location convergence-limiting singularity \textit{a priori}. Still, we shall show that, as in the problem above, precise knowledge of this singularity is not required to obtain a convergent expansion.   

\section{Meniscus at a flat wall \label{sec:Meniscus}}
\subsection{Background and formulation \label{sec:MeniscusBackground}}

To examine the nature of the exponential gauge function transformation used in section~\ref{sec:SakiadisMapped}, we examine a second classical problem which has similar structural features.  In particular, we consider the well-studied problem of the shape of a static meniscus rising above an infinite horizontal pool~\citep{Batchelor,Probstein}.   The pool of liquid has density, $\rho$, is subjected to gravity, $g$, and is in contact with air of negligible density.  A flat wall is placed vertically in the pool, and the liquid intersects the wall with a contact angle $\theta$ (measured through the liquid) as shown in figure~\ref{fig:ErrorPlot}b; for purposes of this study, we assume that $\theta$ lies between $0$ and $\pi/2$. The location of the air--liquid interface, with surface tension $\sigma$, is parameterized as $y=h(x)$ where $x$ is the horizontal distance from the wall, and $y=0$ is the undisturbed location of the interface as $x\to\infty$.  For this configuration, the Young--Laplace equation couples with the hydrostatic field to yield the following dimensionless equations and boundary conditions~\citep{Batchelor,Probstein}: 
\begin{subequations}
\begin{equation}
    \bar{h}= \frac{\bar{h}''}{\left[1+\left(\bar{h}'\right)^{2}\right]^{3/2}},
     \label{eq:secondorderODE}
\end{equation}
\begin{equation}
    \bar{h}'(0)= -\cot\theta,
    \label{eq:slopeboundarycondition}
\end{equation}
\begin{equation}
\bar{h}(\infty)=0.
\label{eq:flatcondition}
\end{equation}
\label{eq:secondorder}
\end{subequations}
In~(\ref{eq:secondorder}), derivatives in $\bar{x}$ are denoted with primes, and the over-bars denote dimensionless variables defined as
\begin{equation}
    \bar{h}=\frac{h}{L},~\bar{x}=\frac{x}{L},~ L=\sqrt{\frac{\sigma}{\rho g}}.
    \label{eq:dlvars}
\end{equation}
In~(\ref{eq:dlvars}), the characteristic length scale $L$ is the well-known capillary length. Multiplying both sides of~(\ref{eq:secondorderODE}) by $\bar{h}'$, integrating, and applying the boundary conditions~(\ref{eq:slopeboundarycondition}) and~(\ref{eq:flatcondition}), we obtain 
\begin{subequations}
\begin{equation}
    \bar{h}'= - \left[\frac{1}{\left(1-\frac{1}{2}\bar{h}^{2}\right)^{2}}-1\right]^{1/2}
    \label{eq:firstorderOperator}
\end{equation}
with the constraint
\begin{equation}
    \bar{h}(0)=\sqrt{2(1-\sin\theta)},
    \label{eq:heightatthewall}
\end{equation}
\label{eq:firstorderODE}
\end{subequations} 
where~(\ref{eq:heightatthewall}) represents the height of the interface at the wall as function of the contact angle, $\theta\in[0,\pi/2]$. Note that, in~(\ref{eq:firstorderOperator}), $\bar{h}'\to-\infty$ as $\bar{h}\to\sqrt{2}$. 

The exact inverse solution $\bar{x}(\bar{h})$ of ~(\ref{eq:firstorderODE}) is obtainable via variable separation and integration~\citep{Batchelor} and is given as
\begin{equation}
    \bar{x}= \cosh^{-1}\frac{2}{\bar{h}} - \cosh^{-1}\sqrt{\frac{2}{1-\sin\theta}} + \sqrt{2+2\sin\theta} - \sqrt{4-{\bar{h} ^{2}}}.
    \label{eq:Batcher}
\end{equation}
In what follows, we obtain an analytic solution for $\bar{h}(\bar{x})$ directly via series expansion.  Our intention in doing so is not to replace equation~(\ref{eq:Batcher}) in usage, but it is to elucidate structural similarities and provide insights to the Sakiadis series solution in section~(\ref{sec:SakiadisMapped}).  Equation~(\ref{eq:Batcher}) is used in what follows to assess the accuracy of the series solutions provided in sections~\ref{sec:divergent} and~\ref{sec:MeniscusMapped}. 

\subsection{Divergent power series solution \label{sec:divergent}}
The standard power series solution of~(\ref{eq:firstorderODE}) is found by assuming
\begin{subequations}
\begin{equation}
    \bar{h}=\sum_{n=0}^\infty \alpha_{n}\bar{x}^{n},~~|\bar{x}|<|\bar{x}_s(\theta)|,
    \label{eq:meniscusform}
\end{equation}
substituting~(\ref{eq:meniscusform}) into~(\ref{eq:firstorderODE}), using Cauchy's product rule~\citep{Churchill} to evaluate $\bar{h}^2$, and applying JCP Miller's formula~\citep{Henrici} for raising a series to a power to obtain the following recursions for the coefficients:
\begin{eqnarray}
    &\alpha_{n+1}=\frac{-2}{n+1}d_{n},~\alpha_0=\sqrt{2(1-\sin\theta)}\\
         &d_{n>0}= \displaystyle\frac{1}{n\widetilde{c}_{0}}\sum_{j=1}^n\left(\frac{3}{2}j-n\right)\widetilde{c}_{j}d_{n-j},~d_{0}=\widetilde{c}_{0}^{\frac{1}{2}},~\widetilde{c}_{n>0}=c_{n>0},~\widetilde{c}_{0}=c_{0}-\frac{1}{4},\\
              &c_{n>0}= \displaystyle\frac{1}{n\widetilde{b}_{0}}\sum_{j=1}^n(-j-n)\widetilde{b}_{j}c_{n-j},~c_{0}=(\widetilde{b}_{0})^{-2},~\widetilde{b}_{n>0}=b_{n>0},~ \widetilde{b}_{0}=b_{0}-2,\\
                  &b_{n}=\displaystyle\sum_{j=0}^n \alpha_{j}\alpha_{n-j}.
\end{eqnarray}
\label{eq:divpowerseries}
\end{subequations}

In~(\ref{eq:meniscusform}), the series is stated to converge within the region $|\bar{x}|<|\bar{x}_s(\theta)|$ with $\bar{x}_s(\theta)$ being the as-of-yet undetermined closest singularity to $\bar{x}=0$.  Without formal proof, we conjecture that $\bar{x}_s(\theta)$ is the value that satisfies $\bar{h}(\bar{x}_s)=\sqrt{2}$ for a given $\theta$ in the exact solution~(\ref{eq:Batcher}), causing $\bar{h}'\to-\infty$ in~(\ref{eq:secondorderODE}).    The value of the conjectured limiting singularity $\bar{x}_s(\theta)$ is known exactly in closed-form from the substitution of $\bar{h}=\sqrt{2}$ into~(\ref{eq:Batcher}) and is given as
\begin{equation}
    \bar{x}_s(\theta)= \cosh^{-1}\sqrt{2} - \cosh^{-1}\sqrt{\frac{2}{1-\sin\theta}} + \sqrt{2+2\sin\theta} - \sqrt{2},
    \label{eq:BatcherSing}
\end{equation}
where we note that  $\bar{x}_s\leq0$ for $\theta\in[0,\pi/2]$.  We support this conjecture in what follows.

\begin{figure}[h!]%
    \centering
    \subfloat{{\includegraphics[width=9cm]{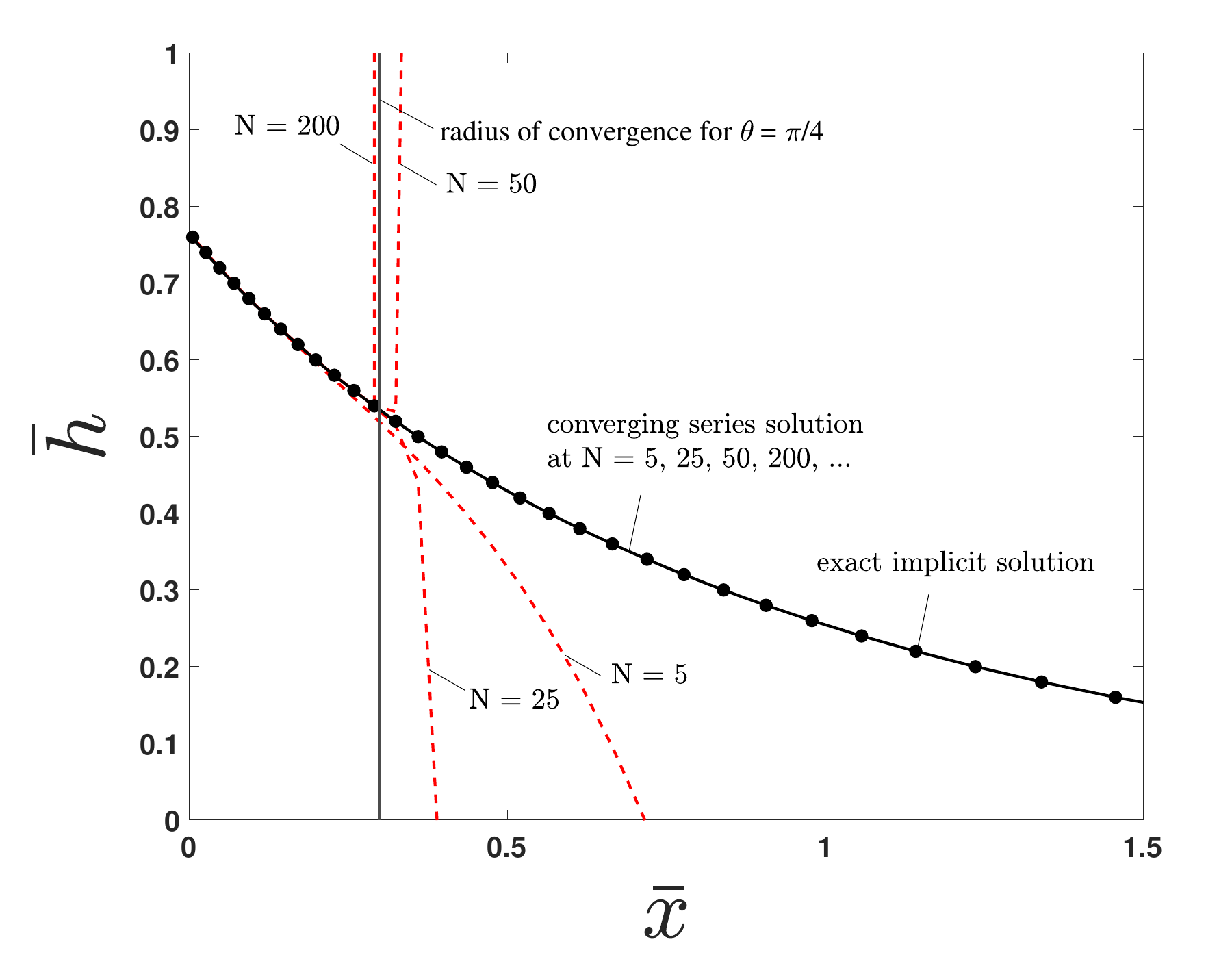} }}%
    \caption{ The solution to~(\ref{eq:firstorderODE}) is shown for a contact angle of $\theta=\pi/4$.  The $N$-term truncations of the divergent series~(\ref{eq:divpowerseries}) (dashed curves) and the convergent resummation~(\ref{eq:TransformedODESolu}) (solid curves) are compared against the exact solution~(\ref{eq:Batcher}) ($\bullet$'s).  The solid vertical line shows the radius of convergence, $|\bar{x}_s(\pi/4)|\approx 0.3$, computed from~(\ref{eq:BatcherSing}).}%
    \label{fig:Interface_45deg}%
\end{figure}

\begin{figure} [h!]
    \centering
    \subfloat{{\includegraphics[width=9cm]{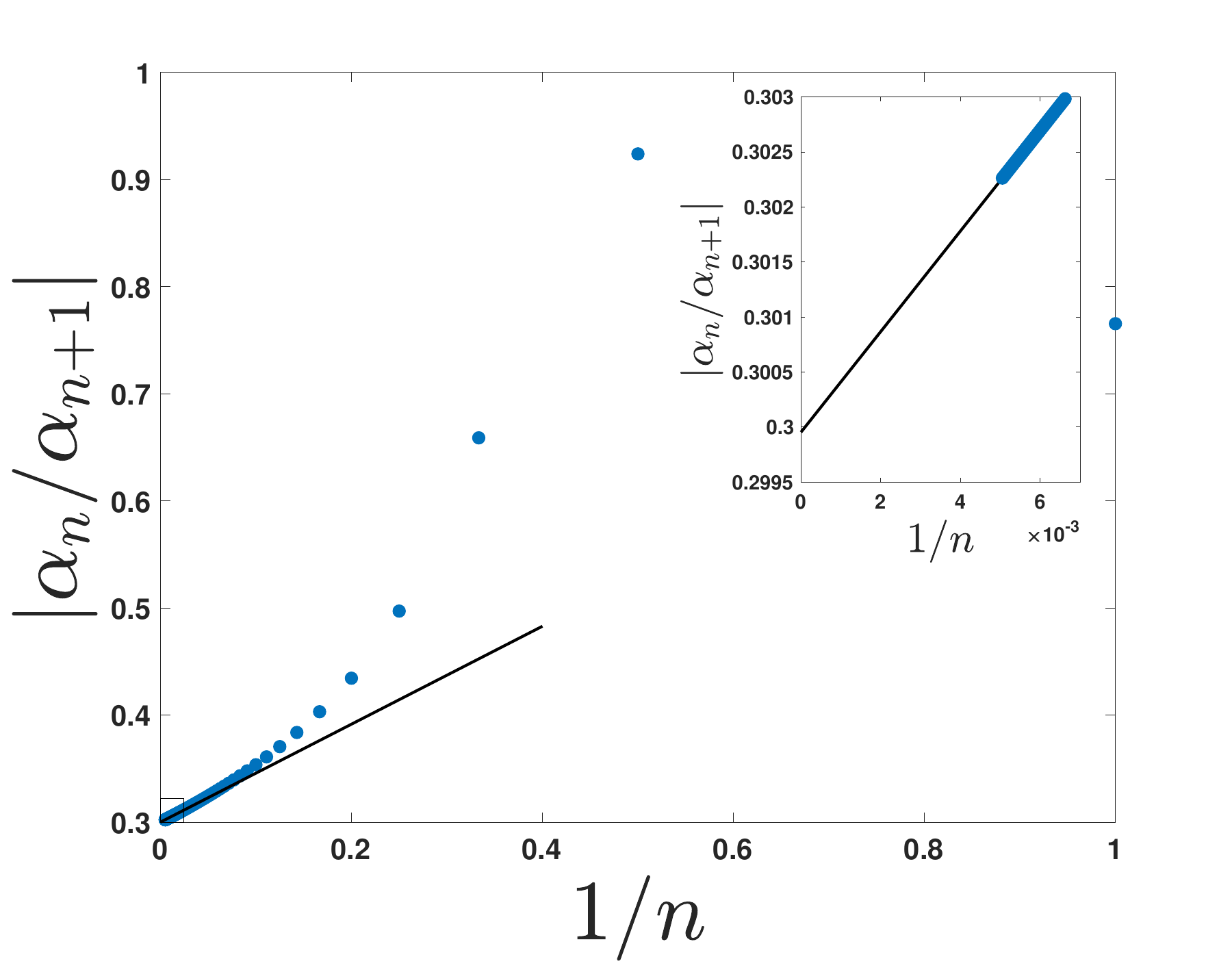} }}%
    \caption{Domb-Sykes plot for~(\ref{eq:divpowerseries}) with $\theta=\pi/4$, indicating (via the ratio-test) a radius of convergence of $|\bar{x}_s(\pi/4)|\approx 0.3$, which is consistent with what is observed in~(\ref{fig:Interface_45deg}).}
    \label{fig:Domb-SkyesPlot_OriginalODE}%
\end{figure}

Figure~(\ref{fig:Interface_45deg}) shows $N$-term truncations of series~(\ref{eq:divpowerseries}) (dashed curves), compared with the exact solution~(\ref{eq:Batcher}) ($\bullet$'s) for $\theta=\pi/4$.  As expected the series~(\ref{eq:divpowerseries}) agrees for small $\bar{x}$ but ultimately diverges at a finite radius of convergence (indicated as a solid vertical line in the figure) and given by $|\bar{x}_s(\pi/4)|\approx 0.3$ from~(\ref{eq:BatcherSing}).
Although we have not formally established that $\bar{x}_s$ is the closest singularity to $\bar{x}=0$ in~(\ref{eq:firstorderODE}) and thus is responsible for divergence, evidence to support this conjecture is given by the Domb-Sykes plot~\citep{DombSykes} in figure~\ref{fig:Domb-SkyesPlot_OriginalODE}, where the magnitude of the ratios of coefficients of~(\ref{eq:divpowerseries}) is shown to approach $|\bar{x}_s(\pi/4)|$ as $n\to\infty$ (i.e., $1/n\to0$ in the figure).  Although only shown in this section for $\theta=\pi/4$, all permissible $\theta$ values lead to similarly divergent series, limited by a radius of convergence of $|\bar{x}_s(\theta)|$.  In section~\ref{sec:MeniscusMapped}, we apply the same type of transformation used in section~\ref{sec:SakiadisMapped} to overcome this convergence barrier for all contact angles. 

\subsection{Asymptotic expansion, variable transform, and convergent power series solution     \label{sec:MeniscusMapped}}

Now that we have a solution to~(\ref{eq:firstorderOperator}), given by~(\ref{eq:divpowerseries}) that—although ultimately divergent beyond some positive $\bar{x}$-value—provides the correct interface shape close to the wall, the asymptotic behavior away from the wall is examined. To this end, the method of dominant balance~\cite{Bender} is employed. To meet the constraint ~(\ref{eq:flatcondition}), it is assumed that
\begin{subequations}
\begin{equation}
    \overline{h}(\overline{x})\ll1,
\end{equation}
and this also assures that its derivatives are small in the limit. To lowest order, then, the governing equation~(\ref{eq:secondorderODE}) is approximated as:
\begin{equation}
    \overline{h}'' \sim\overline{h},~~\overline{x}\to\infty.
\end{equation}
\label{eq:asympformODE}
\end{subequations}
The asymptotic solution to~(\ref{eq:asympformODE}) is
\begin{subequations}
\begin{equation}
    \overline{h} \sim C_0e^{-\overline{x}},~~\overline{x}\to\infty
\end{equation}
where $C_0$ is an arbitrary constant. To obtain the next correction, we assume the following expansion
\begin{equation}
    \overline{h} \sim C_0e^{-\overline{x}}+ D(\overline{x}),~~\overline{x}\to\infty,
    \label{eq:asympform}
\end{equation}
where 
\begin{equation}
    D(\overline{x}) \ll C_0 e^{-\overline{x}},~~\overline{x}\to\infty.
    \label{eq:asympRelation}
\end{equation}
Equation~(\ref{eq:asympform}) is substituted in to ~(\ref{eq:secondorderODE}), subdominant terms are neglected, and the resulting linear equation is solved in accordance with the asymptotic relation~(\ref{eq:asympRelation}) to obtain:
\begin{equation}
    D(\overline{x}) \sim \frac{3}{16}C_0^{3}e^{-3\overline{x}},~~\overline{x}\to\infty.
\end{equation}
\end{subequations}
The same process is repeated to generate higher order correction to~(\ref{eq:asympform}). We obtain:
\begin{equation}
    \overline{h}\sim C_0e^{-\overline{x}} + \frac{3}{16}C_0^{3}e^{-3\overline{x}} +\frac{3}{256}C_0^{5}e^{-5\overline{x}} + O(e^{-7\overline{x}}),~~\overline{x}\to\infty.
    \label{eq:asympformFinal}
\end{equation}
The pattern of exponentials is thus evident in equation~(\ref{eq:asympformFinal}). We note here that a similar exponential behavior is also observed in the Sakiadis problem previously and provides a linkage between the problems in approach and interpretation to follow.


The asymptotic solution~(\ref{eq:asympformFinal}) motivates us to transform ~(\ref{eq:firstorderODE}) to reflect the exponential pattern of~(\ref{eq:asympformFinal}).  This is achieved by  transformations in both the independent and dependent variables, given respectively as
\begin{subequations}
\begin{equation}
     U(\bar{x})=e^{-2\bar{x}},
     \label{eq:indtrans}
\end{equation}
\begin{equation}
   H\left(U(\bar{x})\right)=\bar{h}(\bar{x})e^{\bar{x}}.
\end{equation}
\label{eq:transformations}
\end{subequations}
Substituting~(\ref{eq:transformations}) into~(\ref{eq:firstorderODE}) leads to the transformed ODE:
\begin{equation}
    \left[H+2U\dot{H}\right]\left[UH^{2}-2\right] = -H\sqrt{4-UH^{2}},~~
    \label{eq:transformedODE}
\end{equation}
with transformed condition
\begin{equation}
    H(1)=\sqrt{2(1-\sin\theta)},
    \label{eq:onecondition}
\end{equation}
where $\dot{H}$ denotes the derivative with respect to $U$.  Although an exact explicit solution to~(\ref{eq:transformedODE}) cannot be found as $H(U)$, an exact implicit solution in $H$ and $U$ can be found by separating variables,  integrating~(\ref{eq:transformedODE}), and applying~(\ref{eq:onecondition}) to arrive at
\begin{equation}
\sqrt{2+2\sin\theta}-\sqrt{4-UH^2}=\ln\frac{H\left[1+\sqrt{1+\sin\theta}\right]}{\sqrt{1-\sin\theta}\left[2+\sqrt{4-UH^2}\right]}.    \label{eq:impsoln}
\end{equation}
Equation~(\ref{eq:impsoln}) is used to extract the condition
\begin{equation}
H(0)=\frac{4\sqrt{1-\sin\theta }~ e^{-2+\sqrt{2+2\sin\theta}}}{\sqrt{1+\sin\theta}+\sqrt{2}},    \label{eq:zerocondition}
\end{equation}
which implies $H(0) = C_0$ in the asymptotic solution~(\ref{eq:asympformFinal}). This is needed for the series solution that follows. 

Using the same procedure as in sections~(\ref{sec:SakiadisMapped}) and~(\ref{sec:divergent}), the power series solution of~(\ref{eq:transformedODE}) (with condition~(\ref{eq:zerocondition})) is given by
\begin{subequations}
\begin{equation}
    H=\sum_{n=0}^\infty A_{n} U^{n},
    \label{eq:InitialTry}
\end{equation}
where,
\begin{equation}
    A_{n>0}=\frac{-\displaystyle\sum_{j=0}^{n-1} A_{j} [(1+2j)\mathcal{C}_{n-j}+ \mathcal{D}_{n-j}]}{\mathcal{D}_{0}+\mathcal{C}_{0}(1+2n)},~~A_{0}=\frac{4\sqrt{1-\sin\theta }~ e^{-2+\sqrt{2+2\sin\theta}}}{\sqrt{1+\sin\theta}+\sqrt{2}},
\end{equation}
\begin{equation}
 \mathcal{D}_{n>0}=\frac{1}{n\widetilde{\mathcal{B}}_{0}}\sum_{j=1}^n (\frac{3}{2}j-n)\widetilde{\mathcal{B}}_{j}{\mathcal{D}}_{n-j},~~
     \mathcal{D}_{0}=\widetilde{\mathcal{B}}_{0}^{\frac{1}{2}},~~\mathcal{C}_{n>0}=\mathcal{B}_{n-1},~ ~\mathcal{C}_{0}=-2
\end{equation}
\begin{equation}
 \widetilde{\mathcal{B}}_{n>0}=-\mathcal{B}_{n-1},~~\widetilde{\mathcal{B}}_{0}=4,~~   \mathcal{B}_{n}=\sum_{j=0}^n A_{j}A_{n-j}.
\end{equation}
Transforming back to $\bar{h}(\bar{x})$ space via~(\ref{eq:transformations}), our expansion about $\bar{x}=\infty$ (i.e., $U$=0) is
\begin{equation}
    \bar{h}=e^{-x}\sum_{n=1}^\infty A_{n}\left(e^{-2\bar{x}}\right)^n,
    \label{eq:transformedSeries}
\end{equation}
\label{eq:TransformedODESolu}
\end{subequations}
which, by construction, is consistent with the asymptotic ordering~(\ref{eq:asympformFinal}) as $\bar{x}\to\infty$, and shows explicitly the exponential gauge function $e^{-2\bar{x}}$.

\begin{figure} [h!]
    \centering
    \subfloat{{\includegraphics[width=8cm]{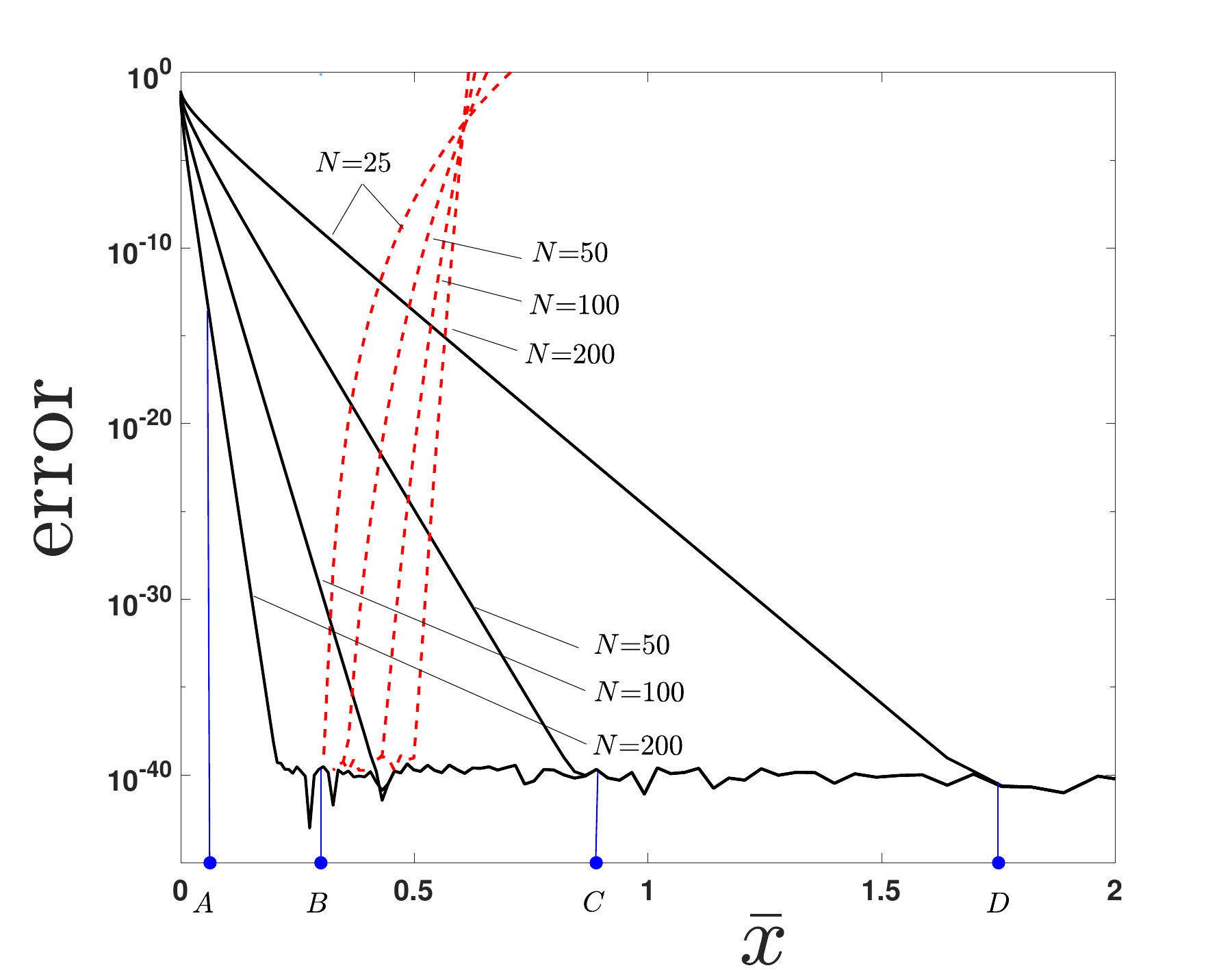} }}
    \subfloat{\hspace{-0.2in}{\includegraphics[width=8cm]{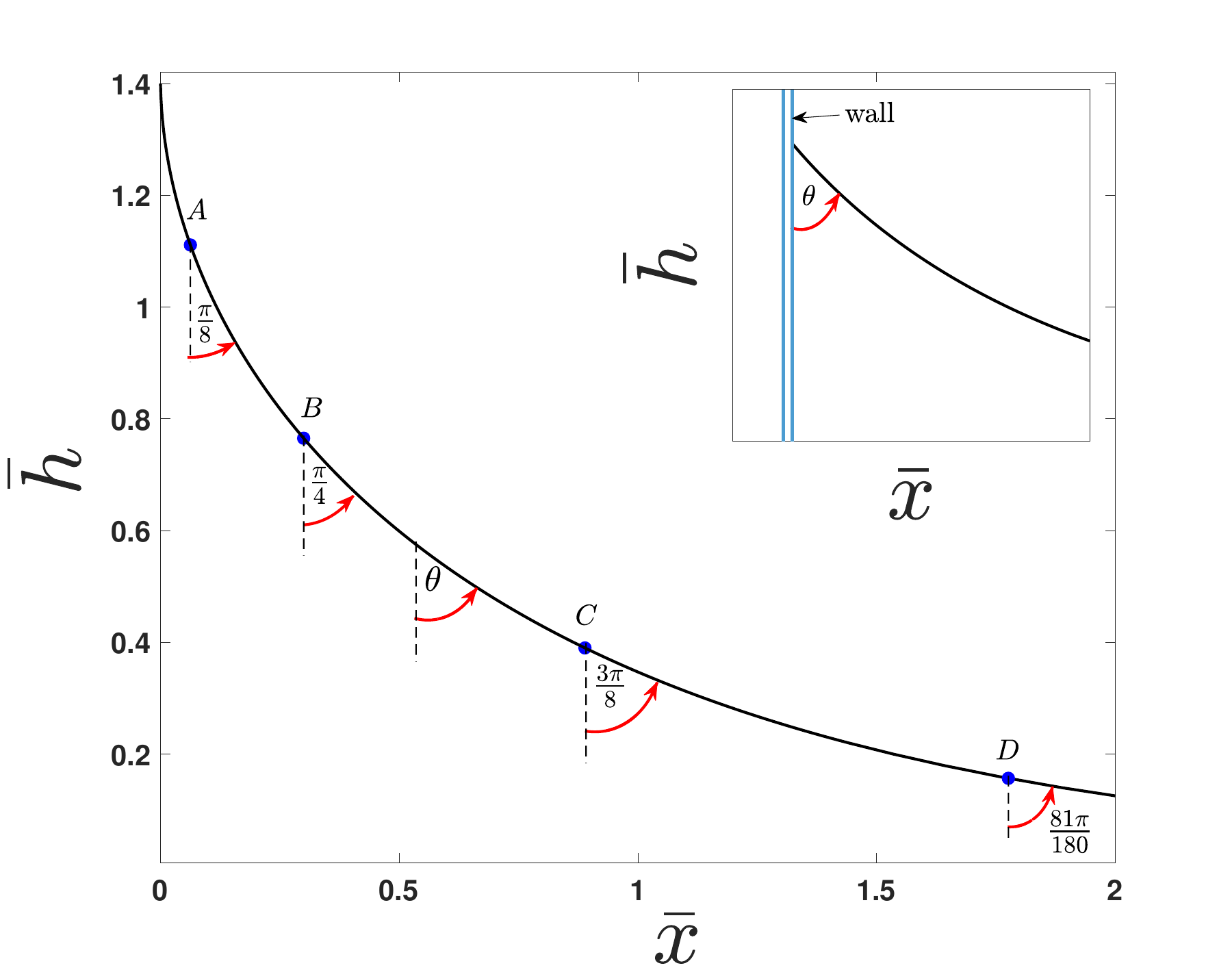} }}%
    \caption{(a) Absolute error between $N$-term truncations of series~(\ref{eq:TransformedODESolu}) (solid lines) and the exact solution to~(\ref{eq:firstorderODE}) given by~(\ref{eq:Batcher}) for $\theta=0$.  Labeled vertical lines show the wall at various contact angles specified in the adjacent figure. For $\theta=\pi/4$, the wall is at point $B$ and the error between the divergent series~(\ref{eq:divpowerseries}) and~(\ref{eq:Batcher}) is shown (\textcolor{red}{dashed line}).  (b) ``Master'' solution to~(\ref{eq:firstorderODE}) at $\theta = 0$, illustrating the $\theta$ values at points $A$, $B$, $C$, and $D$ for potential wall locations in (a).}
    \label{fig:ErrorPlot}%
\end{figure}

\begin{figure} [h!]
\centering
     \subfloat{{\includegraphics[width=9cm]{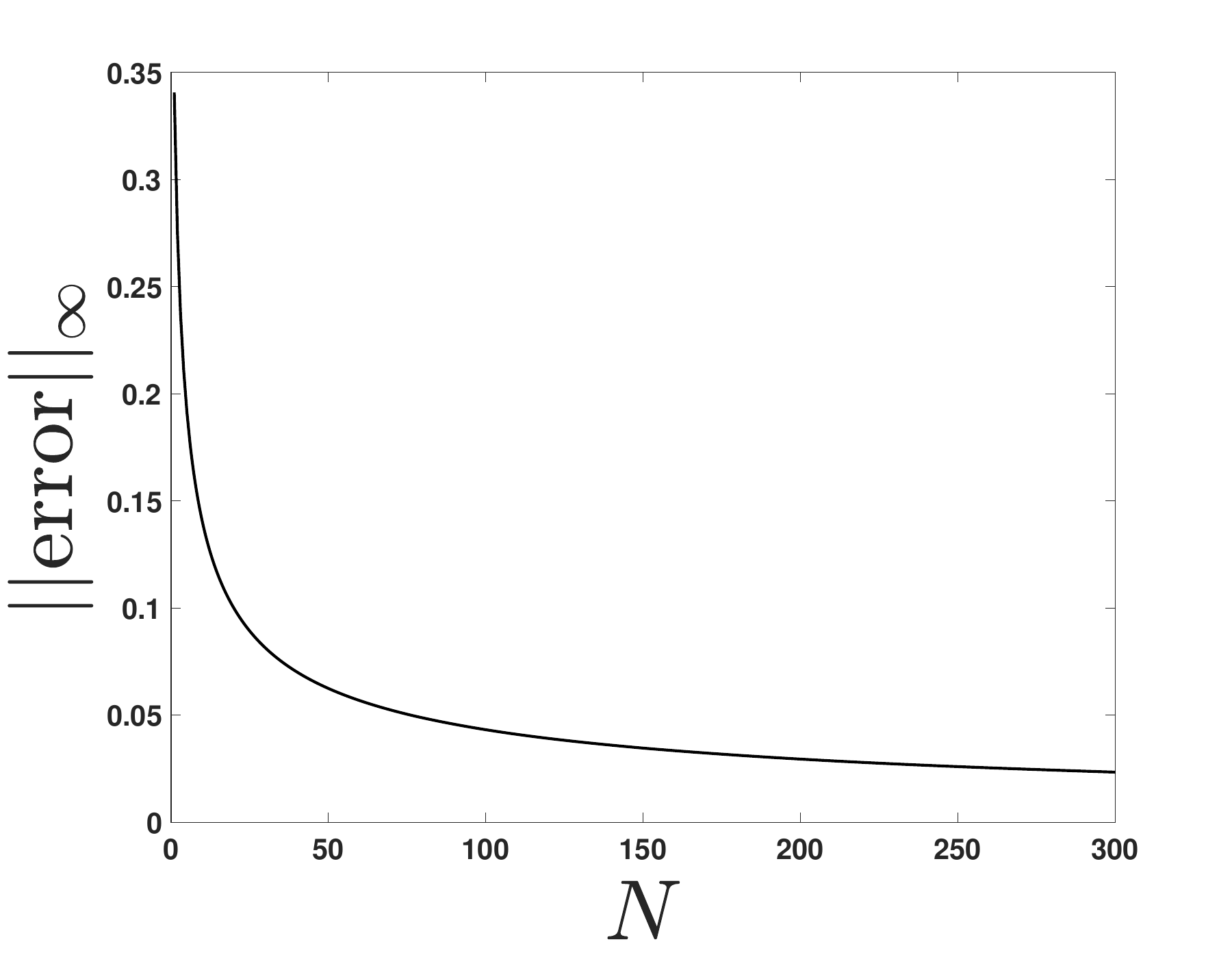} }}%
    \caption{Maximum absolute error between $N$-terms truncations of (\ref{eq:TransformedODESolu}) and~(\ref{eq:Batcher}) (occurring at $\bar{x}=0$ and $\theta=0$), plotted versus $N$.}
    \label{fig:ErrorvsN}
\end{figure}

Figure~\ref{fig:Interface_45deg} shows $N$-term truncations of~(\ref{eq:TransformedODESolu}) (solid curves) compared with both the exact solution~(\ref{eq:Batcher}) and divergent series~(\ref{eq:divpowerseries}). The difference between~(\ref{eq:TransformedODESolu}) and the exact solution is not discernible on the scale of the figure for any $N$ shown, and it is noteworthy to mention that the radius of convergence of the original series~(\ref{eq:divpowerseries}) has been exceeded.  The absolute error between~(\ref{eq:TransformedODESolu}) and the exact solution~(\ref{eq:Batcher}) is shown in figure~\ref{fig:ErrorPlot}a, which indicates convergence (as $N$ increases) for a continuum of angles $\theta$, as prescribed in figure~\ref{fig:ErrorPlot}b.  Figure~\ref{fig:ErrorPlot} is generated for the smallest possible contact angle ($\theta=0$) and thus, by virtue of~(\ref{eq:firstorderODE}) being an autonomous ODE, contains interface shapes for all contact angles as shifted semi-infinite domains; this is indicated in the figure. The maximum error occurs at the wall and is shown versus $N$ for $\theta=0$ in figure~\ref{fig:ErrorvsN}.  

\begin{figure} [h!]
    \centering
    \subfloat{{\includegraphics[width=7.8cm]{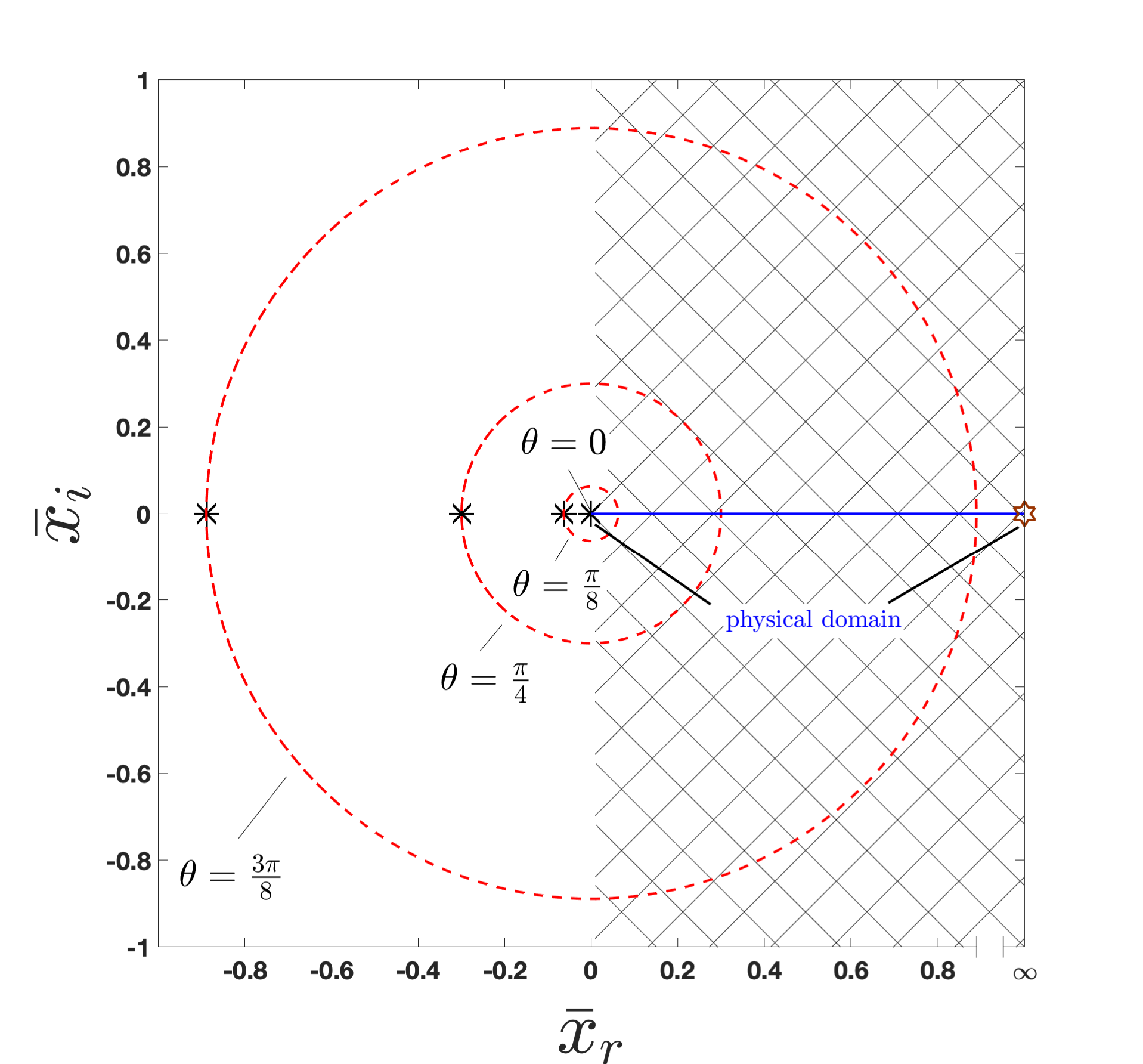} }}%
    \subfloat{{\includegraphics[width=7.8cm]{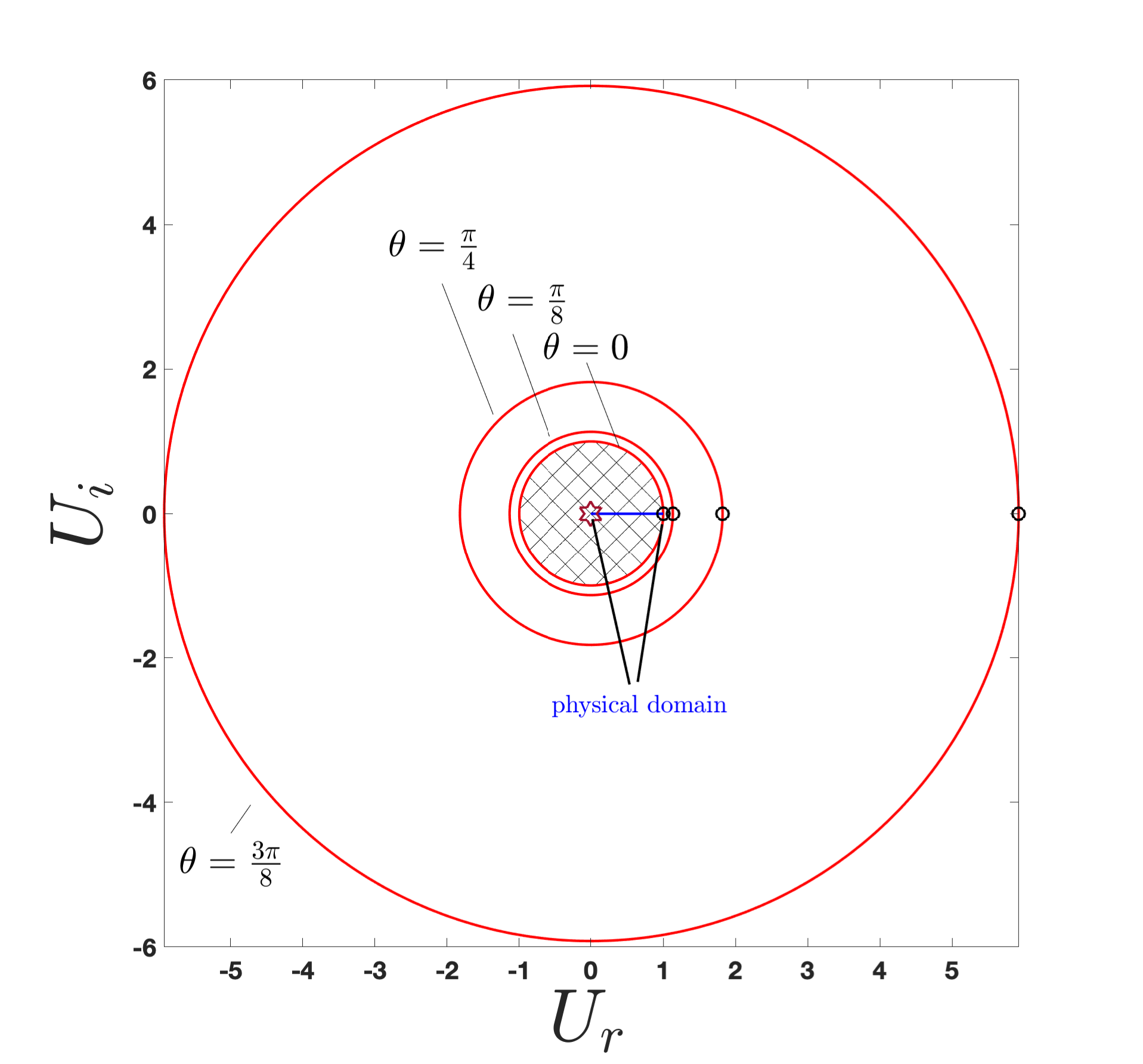} }}%
    \caption{The effect of the gauge transformation~(\ref{eq:indtrans}) on the singularities $\bar{x}_s$ given by~(\ref{eq:BatcherSing}) for various $\theta$ values, showing their placement on the negative real $\bar{x}$ line ($\ast$, left) and their image on the positive real $U$ line ($\circ$, right). The \textcolor{red}{dashed circles} shown in the $\bar{x}$ plane deliniate the regions of convergence of series~(\ref{eq:divpowerseries}) for various indicated $\theta$ values.  The circles shown in the $U$ plane delineate the regions of convergence of~(\ref{eq:TransformedODESolu}) for the same $\theta$ values.}%
    \label{fig:MeniscusMapped}%
\end{figure}

\begin{figure} [h!]
    \centering
    \subfloat{{\includegraphics[width=9cm]{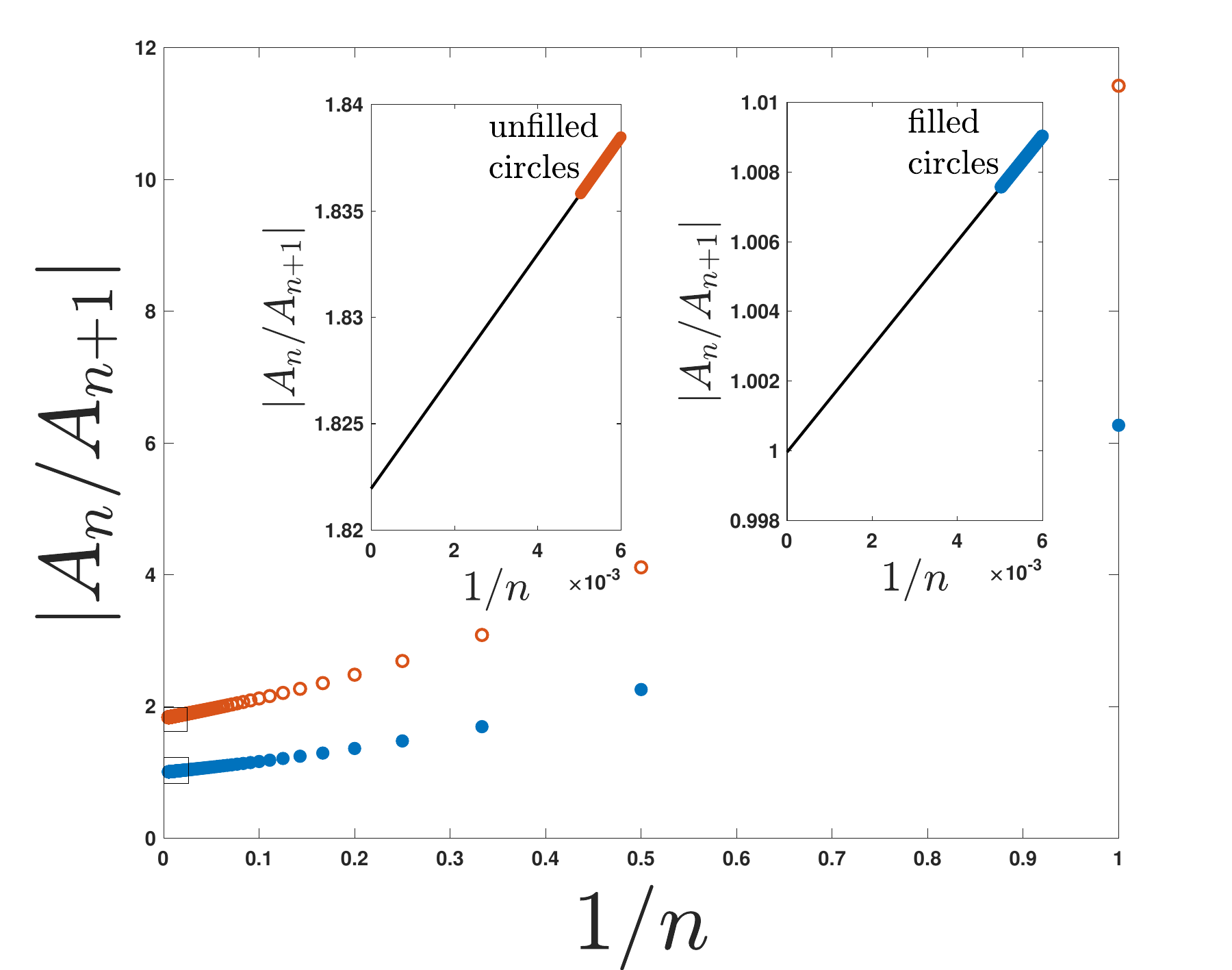} }}%
    \caption{Domb-Sykes plot for~(\ref{eq:TransformedODESolu}), for $\theta = \pi/4$ (upper curve, $\circ$) and $\theta = 0$ (lower curve, $\bullet$), indicating radii of convergence consistent with figure~\ref{fig:MeniscusMapped}.}
    \label{fig:Domb-SkyesPlot_TransformedODE}%
\end{figure}

As done in~(\ref{sec:SakiadisMapped}) for the Sakiadis problem, we now provide an explanation of why the series~(\ref{eq:TransformedODESolu}) converges to the exact solution~(\ref{eq:Batcher}) of the ODE~(\ref{eq:firstorderODE}) describing a meniscus at a flat wall.  The answer again lies in the mapping provided by the gauge function (here,~(\ref{eq:indtrans})), as shown in figure~\ref{fig:MeniscusMapped} where the complex $\bar{x}$ and $U$ planes are compared. In the $\bar{x}$ plane of figure~\ref{fig:MeniscusMapped}, circles of convergence for~(\ref{eq:divpowerseries}) centered around $\bar{x}$=0 are drawn for various $\theta$ values, based on the conjectured closest singularity (to $\bar{x}$=0), $\bar{x}_s(\theta)$, using~(\ref{eq:BatcherSing}). For $\theta=0$, no circle is drawn, since $\bar{x}_s(\theta)$ (which sets the radius) is 0.  Note, that these circles of convergence each intersect the positive real line.  In particular, for $\theta=\pi/4$, this intersection occurs at the same location as the radius of convergence of series~(\ref{eq:divpowerseries}) drawn in figure~\ref{fig:Interface_45deg}; this correspondence holds for all other values of $\theta$, as expected from Taylor's theorem. In the $U$ plane of figure~\ref{fig:MeniscusMapped}, the circle of convergence for~(\ref{eq:TransformedODESolu}) is drawn, centered at $U=0$ and with mapped radius of $|e^{-2\bar{x}_s}|$; note that this circle now extends to the boundary of the physical domain for $\theta=0$ and extends beyond it for $\theta>0$, which explains why~(\ref{eq:TransformedODESolu}) converges over the entire physical domain for all values of $\theta$.  
The circles of convergence in figure~\ref{fig:MeniscusMapped} are, of course, still \textit{conjectured} because, in constructing the mapping in figure~\ref{fig:MeniscusMapped}, we are assuming that the singularity $\bar{x}_s(\theta)$ (given by~(\ref{eq:BatcherSing})) is the closest singularity to $\bar{x}=0$ and that no other singularities map to $U$ singularities closer to $U=0$.  In addition to evidence given in section~\ref{sec:divergent}, further evidence that supports this conjecture is provided in the Domb-Sykes plots in figure~\ref{fig:Domb-SkyesPlot_TransformedODE}, where the radius of convergence deduced from this numerical ratio-test is consistent with the locations of the mapped singularities in figure~\ref{fig:MeniscusMapped}. 

Finally, it is worth noting that---like in the Sakiadis problem---knowledge of these singularities is not required to implement the exponential gauge function transformation~(\ref{eq:transformations}) that leads to a convergent series solution.  

\section{Conclusions \label{sec:conclusions}}
In this work, we provide convergent power series solutions to the Sakiadis boundary layer problem and the problem of a meniscus at a flat wall, by means of transforming the original ODEs in terms of variable substitutions that are motivated by the asymptotic expansions about $\infty$. In both cases, the transformations map the dominant convergence-limiting singularities out of the physical domain; also in both cases,  convergence-limiting singularities do not need to be known \textit{a priori}  but their locations are deduced nonetheless.  For the Sakiadis problem, this provides---in the absence of a formal proof of exactness---a \textit{conjectured} exact Taylor series representation of the solution over the full physical domain. That said, for both the Sakiadis and meniscus problems, the exponential gauge functions used handle singularities similarly to achieve these demonstrably convergent series solutions.  

Although the nature of nonlinear ODEs precludes general conclusions, our results indicate that asymptotic behaviors can be useful to motivate gauge functions to overcome power series divergence.  Additionally, the approach used here supports a growing body of literature~\citep{Barlow:2017,Barlow:2017b,Beachley,Belden,SIR2020,SEIR2020,Rame} underscoring the use of power series solutions as a viable method for analytically solving nonlinear ODEs.

\bibliographystyle{imamat}
\bibliography{meniscussakiadis}

\end{document}